\documentclass[letterpaper,aps,prb,twocolumn,superscriptaddress,amsmath,amssymb,floatfix]{revtex4}
\pdfoutput=1
\usepackage{lmodern}
\usepackage{graphicx}
\usepackage{bm}
\usepackage{color}
\usepackage[percent]{overpic}
\usepackage[%
breaklinks=true,
colorlinks,citecolor=blue,linkcolor=blue,urlcolor=blue,pdftitle={Non-equilibrium dynamics of photo-excited electrons in graphene: collinear scattering, Auger processes, and the impact of screening},pdfauthor={A. Tomadin, D. Brida, G. Cerullo, A.C. Ferrari, and M. Polini},pdfsubject={},pdfcreator={},pdfproducer={},pdfkeywords={}]{hyperref}
\usepackage[normalem]{ulem}
\begin{document}
\title{Non-equilibrium dynamics of photo-excited electrons in graphene: collinear scattering, Auger processes, and the impact of screening}
\author{Andrea Tomadin}
\email{andrea.tomadin@sns.it}
\affiliation{NEST, Istituto Nanoscienze-CNR and Scuola Normale Superiore, I-56126 Pisa, Italy}
\author{Daniele Brida}
\affiliation{Department of Physics and Center for Applied Photonics, University of Konstanz, D-78457 Konstanz, Germany}
\affiliation{IFN-CNR, Dipartimento di Fisica, Politecnico di Milano, P.za Leonardo da Vinci, 20133 Milano, Italy}
\author{Giulio Cerullo}
\affiliation{IFN-CNR, Dipartimento di Fisica, Politecnico di Milano, P.za Leonardo da Vinci, 20133 Milano, Italy}
\author{Andrea C. Ferrari}
\affiliation{Cambridge Graphene Centre, University of Cambridge, 9 JJ Thomson Avenue, Cambridge, CB3 OFA, UK}
\author{Marco Polini}
\affiliation{NEST, Istituto Nanoscienze-CNR and Scuola Normale Superiore, I-56126 Pisa, Italy}

\begin{abstract}
We present a combined analytical and numerical study of the early stages (sub-$100~{\rm fs}$) of the non-equilibrium dynamics of photo-excited electrons in graphene.
We employ the semiclassical Boltzmann equation with a collision integral that includes contributions from electron-electron (e-e) and electron-optical phonon interactions.
Taking advantage of circular symmetry and employing the massless Dirac Fermion (MDF) Hamiltonian, we are able to perform an essentially analytical study of the e-e contribution to the collision integral.
This allows us to take particular care of subtle collinear scattering processes---processes in which incoming and outgoing momenta of the scattering particles lie on the same line---including carrier multiplication (CM) and Auger recombination (AR).
These processes have a vanishing phase space for two dimensional MDF bare bands.
However, we argue that electron-lifetime effects, seen in experiments based on angle-resolved photoemission spectroscopy, provide a natural pathway to regularize this pathology, yielding a finite contribution due to CM and AR to the Coulomb collision integral.
Finally, we discuss in detail the role of physics beyond the Fermi golden rule by including screening in the matrix element of the Coulomb interaction at the level of the Random Phase Approximation (RPA), focusing in particular on the consequences of various approximations including static RPA screening, which maximizes the impact of CM and AR processes, and dynamical RPA screening, which completely suppresses them.
\end{abstract}
\maketitle

\section{Introduction}
\label{sec:Introduction}
Graphene, a two-dimensional (2d) crystal of carbon atoms tightly packed in a honeycomb lattice, is at the center of an ever growing research effort, due to its potential as a platform material for a variety of applications in fields ranging from electronics, to food packaging~\cite{geim_naturemater_2007,castroneto_rmp_2009,peres_rmp_2010,dassarma_rmp_2011,kotov_rmp_2012,charlier_tap_2008, bonaccorso_matertoday_2012}.
In particular, in optoelectronics, photonics, and plasmonics graphene has decisive advantages, such as wavelength-independent absorption, tunability via electrostatic doping, large charge-carrier concentrations, low dissipation rates, high mobility, and the ability to confine electromagnetic energy to unprecedented small volumes~\cite{bonaccorso_naturephoton_2010,koppens_nanolett_2011,grigorenko_naturephoton_2012,brida_arxiv_2012,jens_arxiv_2013}.
These unique properties make it an ideal material for a variety of photonic applications~\cite{bonaccorso_naturephoton_2010}, including fast photodetectors~\cite{xia_naturenanotech_2009,vicarelli_naturemater_2012}, transparent electrodes in displays and photovoltaic modules~\cite{bonaccorso_naturephoton_2010,bae_naturenanotech_2010}, optical modulators~\cite{liu_nature_2011}, plasmonic devices~\cite{echtermeyer_naturecommun_2011,grigorenko_naturephoton_2012}, microcavities~\cite{engel_naturecommun_2012}, ultrafast lasers~\cite{sun_acsnano_2010}, just to cite a few.
Therefore, understanding the microscopic interactions between light and matter is an essential requirement to progress these emerging research areas into technological applications.

When light arrives on a graphene sample it creates a highly non-equilibrium ``hot'' electron distribution (HED), which first relaxes on an ultrafast timescale to a thermalized (but still hot) Fermi-Dirac (FD) distribution and then slowly cools, {\it via} optical and acoustic phonon emission, eventually reaching thermal equilibrium with the lattice.
Pump-probe spectroscopy is a very effective tool to study the non-equilibrium dynamics of hot carriers and has been extensively applied to a variety of graphene samples and other carbon-based materials~\cite{sun_prl_2008,dawlaty_apl_2008,breusing_prl_2009,choi_apl_2009,plochocka_prb_2009,newson_opticsexpress_2009,lui_prl_2010,huang_nanolett_2010,ruzicka_apl_2010,strait_nanolett_2011,hale_prb_2011,obraztsov_nanolett_2011,winnerl_prl_2011,carbone_chemphyslett_2011,kaniyankandy_jphyschemc_2011,breusing_prb_2011,brida_arxiv_2012,tielrooij_arxiv_2012,winnerl_jpcm_2013}.
There is consensus in the literature on the fact that the time scales of the thermalization process, primarily controlled by electron-electron (e-e) interactions, are extremely short, of the order of tens of femtoseconds.
Indeed, early theoretical calculations~\cite{gonzalez_prl_1996,gonzalez_prl_1999,hwang_prb_2007,polini_prb_2008} based on the {\it equilibrium} many-body diagrammatic perturbation theory for an interacting system of massless Dirac Fermions (MDFs) all pointed to ultrashort e-e inelastic carrier lifetimes, with a sensitive dependence on doping.
\begin{figure}[h!]
\includegraphics[width=\columnwidth]{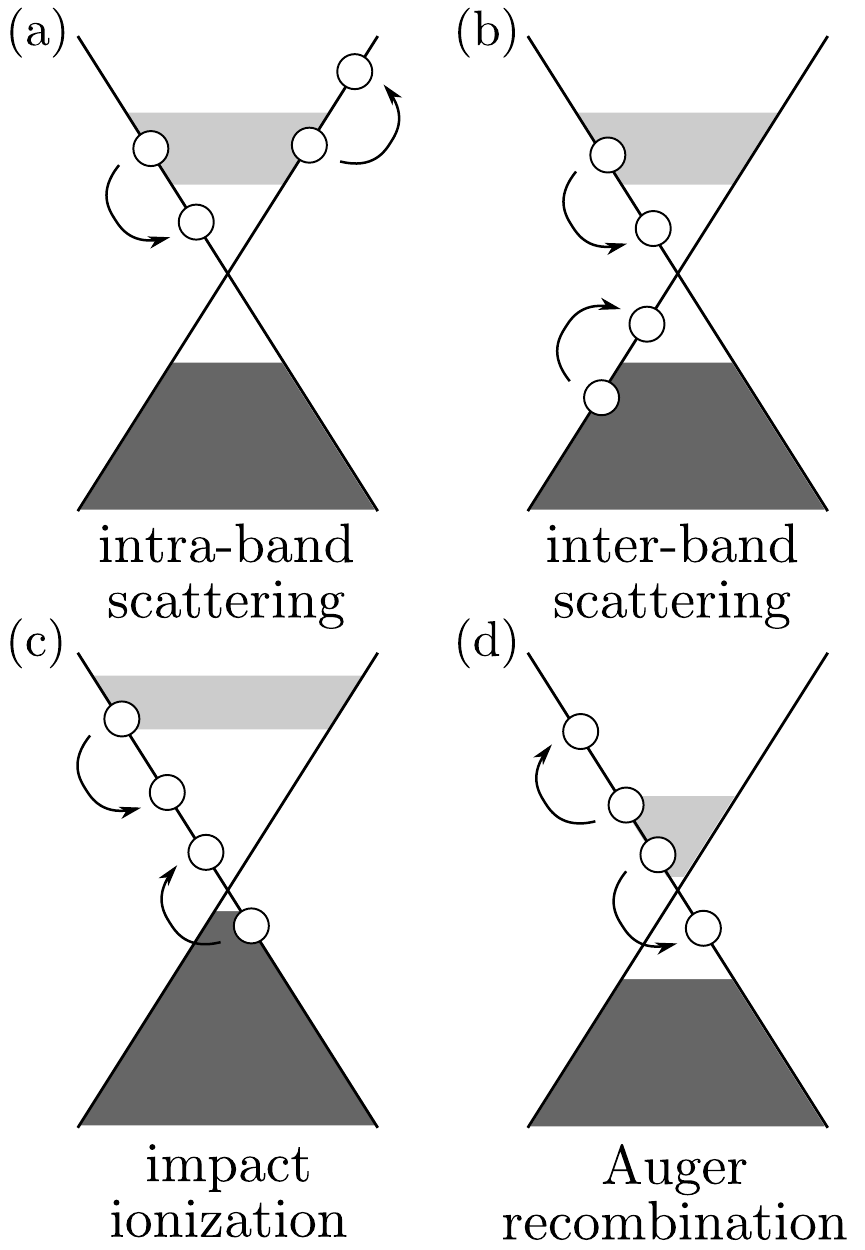}
\caption{Schematic of Coulomb-enabled two-body scattering processes in graphene.
The cones represent the linear dispersion $\varepsilon_{{\bm k},s} = s \hbar v_{\rm F} |{\bm k}|$ of electron states.
Light-gray and dark-gray shaded areas denote occupied states.
These plots correspond to a non-equilibrium hot-electron distribution.
Arrows mark electron transitions from initial to final states.
The electron population in each band is conserved in (a) and (b), but not in (c) and (d).
(c) and (d) represent ``Auger processes,'' which can only take place when the wave vectors of the initial and final states are {\it collinear}.
\label{fig:scatteringprocesses}}
\end{figure}

The theory of the {\it non-equilibrium} dynamics of hot carriers in graphene has also been extensively investigated~\cite{butscher_apl_2007,winzer_nanolett_2010,malic_prb_2011,kim_prb_2011,girdhar_apl_2011,sun_prb_2012,winzer_prb_2012, winzer_jpcm_2013,sun_arxiv_2013}.
Previous works, however, heavily relied on numerical analysis and did not address the following issues.
When electrons in graphene are described by the low-energy 2d MDF model~\cite{castroneto_rmp_2009,peres_rmp_2010,dassarma_rmp_2011,kotov_rmp_2012}, a special class of two-body scattering processes poses a serious conundrum.
These are ``collinear'' events, in which incoming and outgoing momenta of the scattering particles lie on the same line~\cite{rana_prb_2007,kashuba_prb_2008,fritz_prb_2008,schutt_prb_2011} (see Fig.~\ref{fig:scatteringprocesses}).
On one hand, due to the geometrical nature of these events, one is very tempted to conclude that they are irrelevant, since they lie on a one dimensional (1d) manifold embedded in a 2d space, i.e.~a set of zero measure.
As we will see in Sec.~\ref{ssec:coulombkernel}, this intuitive statement can be formally proven by employing conservation of energy and momentum.
Thus, the phase space for collinear scattering events vanishes in the case of 2d MDF bare bands.
On the other hand, when e-e interactions are taken into account going beyond the single-particle picture, several interesting things happen.
i) MDFs moving in a collinear way along the same directrix ``spend a lot of time together'' since they travel with the same speed~\cite{fritz_prb_2008}, the Fermi velocity $v_{\rm F} \sim 10^{6}~{\rm m}/{\rm s}$.
They thus interact very strongly through the non-relativistic Coulomb interaction.
A simple analysis based on the Fermi golden rule shows that this yields~\cite{kashuba_prb_2008,fritz_prb_2008,schutt_prb_2011} logarithmically-divergent quasiparticle decay rates and transport coefficients, such as viscosities and conductivities.
ii) Interactions (even at the Hartree-Fock level~\cite{borghi_ssc_2009}) are responsible for deviations of the energy-momentum dispersion relation from linearity.
The renormalized quasiparticle spectrum, controlled by the real part of the quasiparticle self-energy, displays a concave curvature~\cite{kotov_rmp_2012}, an effect that suppresses collinear scattering.
iii) The broadening of the energy-momentum dispersion, which follows from the finiteness of the quasiparticle lifetime (an effect beyond the Hartree-Fock theory), opens up the phase space for collinear scattering, as thoroughly discussed in Sec.~\ref{sect:Auger}.
The broadening of the quasiparticle spectrum is controlled by the imaginary part of the quasiparticle self-energy, a quantity directly probed by angle-resolved photoemission spectroscopy~\cite{bostwick_naturephys_2007, zhou_naturemater_2007,bostwick_science_2010,walter_prb_2011,siegel_pnas_2011,knox_prb_2011}.
iv) The situation is further complicated by the role of {\it screening}, a key phenomenon in systems with long-range Coulomb interactions~\cite{Pines_and_Nozieres,Giuliani_and_Vignale}.
As we will discuss in Sec.~\ref{sect:screening}, static screening does not have a detrimental effect on collinear scattering.
The opposite occurs when dynamical screening is considered at the level of the Random Phase Approximation (RPA).
v) Non-linearities and anisotropies in the band structure beyond the MDF model (such as ``trigonal warping''~\cite{castroneto_rmp_2009}) may affect the efficiency of screening.
These issues were recently addressed in Ref.~\onlinecite{basko_prb_2013} by means of the equilibrium many-body perturbation theory, as we will discuss in Sec.~\ref{ssec:screeningModels}. 

All these issues raise the following question: is collinear scattering relevant or irrelevant to understand quasiparticle dynamics and transport in graphene?

Collinear (or forward) scattering plays a special role in the dynamics of quasiparticles~\cite{gonzalez_prl_1996} and photo-excited carriers in graphene~\cite{brida_arxiv_2012}.
The finiteness of the quasiparticle lifetime on the mass shell~\cite{gonzalez_prl_1996} can be traced back to the divergence of the density of electron-hole pairs in the collinear direction.
In this case, it is the only configuration in which ``impact ionization'' (IMI) and ``Auger recombination'' (AR) processes are possible~\cite{rana_prb_2007} (see Fig.~\ref{fig:scatteringprocesses}).
IMI and AR (which we will refer to with the generic term ``Auger processes'') have been studied since the later fifties~\cite{BL1959,Keldysh1960}.
In recent years they attracted attention in the context of semiconductors~\cite{LandsbergBook} and quantum dots~\cite{Klimov2000,Nozik2002}.
IMI and AR are of fundamental interest because they strongly influence the relaxation dynamics of a HED.
E.g., AR in optically-pumped 2d electron systems in the quantum Hall regime is responsible~\cite{potemski_prl_1991,plochocka_prb_2009} for emission from states with energy higher than those optically pumped, and thwarts the realization of a Landau-level laser, i.e.~a laser that would operate under the 2d Landau quantization, with population inversion in the Landau levels~\cite{potemski_prl_1991}.
Most importantly, Auger processes can be exploited to design solar cells~\cite{KWWQ1993,WKQ1994} or other photovoltaic devices that can overcome fundamental limitations~\cite{SQ1961} to photocurrent production by relying on ``carrier multiplication'' (CM).

We reported evidence of Auger processes in graphene~\cite{brida_arxiv_2012}, proving the existence of IMI and CM in a short transient following ultrafast photo-excitation in the optical domain~\cite{brida_arxiv_2012}.
The excess energy of photo-excited electrons can also be transferred to secondary electron-hole pairs by intra-band scattering, {\it without} CM from the valence to conduction band.
This process, also recently experimentally demonstrated~\cite{tielrooij_arxiv_2012}, proceeds by promotion of electrons from below to above the Fermi energy and does not involve processes b)-d) in Fig.~\ref{fig:scatteringprocesses}.
On the other hand, Refs.~\onlinecite{cavalleri_arxiv_2013,hofmann_arxiv_2013}, by probing the non-equilibrium dynamics of MDFs by time- and angle-resolved photo-emission spectroscopy, found no evidence for CM.
We note, however, that Refs.~\onlinecite{cavalleri_arxiv_2013,hofmann_arxiv_2013} operated in a regime of pump fluences $\gg 10^{2}~\mu{\rm J}/{\rm cm}^{2}$ where CM is not expected on the basis of calculations relying on static screening~\cite{winzer_prb_2012}.
Moreover, both experiments lacked sufficient time resolution to observe CM.
Indeed, the higher the pump fluence, the shorter is the time window in which CM exists~\cite{winzer_prb_2012}.
E.g., for a pump fluence $\sim 50~{\rm \mu J}/{\rm cm}^2$ as in Ref.~\onlinecite{brida_arxiv_2012}, CM exists in a time window $\sim 100~{\rm fs}$ (substantially larger than the time resolution in Ref.~\onlinecite{brida_arxiv_2012}).
Refs.~\onlinecite{cavalleri_arxiv_2013,hofmann_arxiv_2013} used much higher pump fluences, i.e.~$\gtrsim 1~{\rm mJ}/{\rm cm}^2$ and $\sim 346~{\rm \mu J}/{\rm cm}^2$, respectively.
Ref.~\onlinecite{cavalleri_arxiv_2013} reported strong evidence of population inversion in graphene after intense photoexcitation, similar to what reported in Ref.~\onlinecite{li_prl_2012}, where evidence of stimulated emission was seen for pump fluences $\gtrsim 2~{\rm mJ}/{\rm cm}^2$.
Because of the large fluences in Refs.~\onlinecite{cavalleri_arxiv_2013,li_prl_2012}, the existence of population inversion cannot be ascribed to the absence of Auger processes.

In semiconductors, IMI (AR) creates (annihilates) an electron-hole pair and takes place when the energy transfer to (from) one electron is sufficient to overcome the band gap.
Since graphene is a zero-gap semiconductor, the scattering rates of Auger processes are generally larger than in most other common semiconductors, as discussed in Ref.~\onlinecite{rana_prb_2007}.
However, Ref.~\onlinecite{rana_prb_2007} did not address the issue of the vanishing phase space for 2d MDF bare bands.
Moreover, the IMI and AR rates calculated in Ref.~\onlinecite{rana_prb_2007} refer to FD distributions [Eq.~(20) in Ref.~\onlinecite{rana_prb_2007}], thus do not apply to generic
non-equilibrium situations.
Finally, Ref.~\onlinecite{rana_prb_2007} did not discuss the role of dynamical screening, now known to play a pivotal role in the electronic and optoelectronic properties of graphene~\cite{dassarma_rmp_2011,kotov_rmp_2012}.

Here we analyze in detail the interplay between collinear scattering and e-e interactions in the context of the non-equilibrium dynamics of photo-excited electrons.

We first show that electron lifetime effects open up a finite phase space for collinear scattering processes, thereby regularizing the pathologies mentioned above.
Here we consider the broadening of the energy-momentum dispersion, but we neglect its deviations from linearity due to e-e interactions.
Although these two effects could be treated in principle on an equal footing (since they are described by the imaginary and real part of the quasiparticle self-energy, respectively), changes in the dispersion due to the real part of the quasiparticle self-energy are relevant only for low carrier densities~\cite{kotov_rmp_2012,elias_naturephys_2011}.
While our theory is general, the numerical calculations we present in Sec.~\ref{sec:Results} are focused on a regime with large density of photo-excited carriers, $\sim 10^{13}~{\rm cm}^{-2}$.
This is a value that is typically used in experimental time-resolved techniques for mapping the relaxation dynamics of electron distributions~\cite{brida_arxiv_2012,breusing_prb_2011,hofmann_arxiv_2013}.

We then discuss the contribution of collinear processes to the Coulomb collision integral in the semiclassical Boltzmann equation (SBE), which determines, {\it together} with electron-optical phonon (e-ph) scattering, the early stages (sub-$100~{\rm fs}$) of the time evolution.
Most importantly, we go beyond the Fermi golden rule, by introducing screening at the RPA level.
Contrary to what happens in a conventional 2d parabolic-band electron gas~\cite{giuliani_prb_1982,Giuliani_and_Vignale}, the introduction of {\it dynamical} screening brings in {\it qualitative} new features.
On one hand, RPA dynamical screening represents the most natural and elementary way to regularize~\cite{schutt_prb_2011} the logarithmic divergences of quasiparticle decay rates and transport coefficients~\cite{kashuba_prb_2008,fritz_prb_2008}.
On the other hand, due to a $|\omega^2 - v_{\rm F}^2 q^2|^{-1/2}$ divergence that arises in the polarization function~\cite{polini_prb_2008,wunsch_njp_2006,hwang_prb_2007_lindhard,barlas_prl_2007,principi_prb_2009} $\chi^{(0)}(q,\omega)$ of 2d MDFs when the collinear scattering condition $\omega = \pm v_{\rm F} q$ is met, RPA dynamical screening completely suppresses Auger processes.

This Article is organized as follows.
Sec.~\ref{sec:Hamiltonian} describes the model MDF Hamiltonian and the SBE for the coupled dynamics of electrons and optical phonons.
It also reviews the typical timescales, as set by e-e and e-ph interactions.
Sec.~\ref{sect:collinearscattering} introduces the isotropic SBE and discusses in detail the treatment of collinear scattering in the Coulomb collision integral.
The role of screening is considered in Sec.~\ref{sect:screening}.
Sec.~\ref{sec:Results} presents our main numerical results for the electron and phonon dynamics, as obtained from the solution of the isotropic SBE. Finally, Sec.~\ref{sec:Conclusions}, summarizes our main conclusions.
\section{Model Hamiltonian and the semiclassical Boltzmann equation}
\label{sec:Hamiltonian}

\subsection{MDF Hamiltonian}
\label{sec:MDFHamiltonian}

Carriers in graphene are described in a wide range of energies ($\lesssim 1~{\rm eV}$)
by the MDF Hamiltonian~\cite{castroneto_rmp_2009,peres_rmp_2010,dassarma_rmp_2011,kotov_rmp_2012},
\begin{equation}\label{eq:diagonalkinhamil}
\hat{\cal H}_{\rm MDF} = \sum_{{\bm k}, \ell, \sigma, s}
\varepsilon_{{\bm k},s} \hat{\psi}^{\dag}_{{\bm k}, \ell, s, \sigma}
\hat{\psi}_{{\bm k}, \ell, s, \sigma}~,
\end{equation}
where the field operator $\hat{\psi}_{{\bm k},\ell,s,\sigma}$ annihilates an electron with 2d momentum $\hbar {\bm k}$, valley $\ell = {\rm K}, {\rm K}'$, band index $s = \pm 1$ (or ${\rm c}$, ${\rm v}$ for conduction and valence band, respectively), and spin $\sigma = \uparrow,\downarrow$.
The quantity $\varepsilon_{{\bm k},s} = s \hbar v_{\rm F} |{\bm k}|$ represents the MDF band energy, with a slope $\hbar v_{\rm F} \simeq 0.6~{\rm eV~nm}$.

MDFs interact through the non-relativistic Coulomb potential $v(r) = e^{2} / (\bar{\epsilon} r)$ with the following 2d Fourier transform
\begin{equation}\label{eq:Coulomb}
v_{q} = \frac{2 \pi e^{2}}{\bar{\epsilon} q}~,
\end{equation}
where $\bar{\epsilon} = (\epsilon_1 + \epsilon_2)/2$ is an average dielectric constant~\cite{kotov_rmp_2012} calculated with the dielectric constants $\epsilon_1$ and $\epsilon_2$ of the media above and below the graphene flake.

Intra-valley e-e interactions are described by the following Hamiltonian (in the eigenstate representation):
\begin{eqnarray}\label{eq:interactiontensor}
& & \hat{\cal H}_{\rm e-e} = \frac{1}{2A}\sum_{\ell} \sum_{\sigma_{1},\sigma_{2}} \sum_{\{s_{i}\}_{i=1}^{4}} \sum_{\{{\bm k}_{i}\}_{i=1}^{4}} V^{(\ell)}_{1234} \nonumber \\
& & \times \delta({\bm k}_{1}+{\bm k}_{2} - {\bm k}_{3} - {\bm k}_{4})
\nonumber \\
& & \times \hat{\psi}^{\dag}_{{\bm k}_{1},\ell,s_{1},\sigma_{1}}
\hat{\psi}^{\dag}_{{\bm k}_{2},\ell,s_{2},\sigma_{2}}
\hat{\psi}_{{\bm k}_{4},\ell,s_{4},\sigma_{2}}
\hat{\psi}_{{\bm      k}_{3},\ell,s_{3},\sigma_{1}}~,
\end{eqnarray}
where $A$ is 2d electron system area and the delta distribution imposes momentum conservation.

The matrix element of the Coulomb potential reads
\begin{equation}\label{eq:matrixinteraction}
V^{(\ell)}_{1234} = v_{|{\bm k}_{1} - {\bm k}_{3}|}
F_{s_{1},s_{3}}^{(\ell)}(\theta_{{\bm k}_{3}} - \theta_{{\bm k}_{1}})
F_{s_{2},s_{4}}^{(\ell)}(\theta_{{\bm k}_{4}} - \theta_{{\bm k}_{2}})~,
\end{equation}
where $F_{s_{1},s_{2}}^{(\ell)} (\theta)= \left [ 1 + s_{1} s_{2} \exp{(i \ell \theta )} \right ] / 2$ is the so-called ``chirality factor''~\cite{castroneto_rmp_2009,peres_rmp_2010,dassarma_rmp_2011,kotov_rmp_2012}, which depends on the polar angle $\theta_{{\bm k}_{i}}$ of the wave vector ${\bm k}_{i}$.

The following dimensionless coupling constant~\cite{kotov_rmp_2012} controls the strength of e-e interactions (relative to the typical kinetic energy):
\begin{equation}\label{eq:FSC}
\alpha_{\rm ee} =  \frac{e^2}{\hbar v_{\rm F} \bar{\epsilon}}~.
\end{equation}
\subsection{Electron-electron interactions}

The distribution function $f_{{\bm k},\ell,s,\sigma}$ represents the probability that a given single-particle state with quantum numbers
${\bm k},\ell,s,\sigma$ is occupied.
The equation of motion (EOM) for this distribution function in the presence of e-e interactions is given by~\cite{KadanoffBaym,HaugJauhoBook,snoke_annphys_2011}:
\begin{widetext}
\begin{eqnarray}\label{eq:eescattering}
& & \left . \frac{d f_{{\bm k}_{1}, \ell, s_1, \sigma_1}}{dt} \right |_{\rm e-e}
= \frac{2\pi}{\hbar}~\frac{1}{A^3}\sum_{{\bm k}_2, {\bm k}_3, {\bm k}_4} \sum_{s_2, s_3, s_4}\sum_{\sigma_2} |V^{(\ell,\sigma_{1},\sigma_{2})}_{1234}|^2 \left ( 1 -\frac{\delta_{\sigma_1, \sigma_2}}{2} \right )~ \delta({\bm k}_{1} +{\bm k}_2 - {\bm k}_3 -{\bm k}_4) \nonumber \\
&\times  & \delta(\varepsilon_{{\bm k}_1, s_1} +\varepsilon_{{\bm k}_2, s_2} - \varepsilon_{{\bm k}_3, s_3} - \varepsilon_{{\bm k}_4, s_4}) [(1 - f_{{\bm k}_{1},  \ell, s_1, \sigma_1}) (1- f_{{\bm k}_2, \ell,s_2, \sigma_2})f_{{\bm k}_3, \ell, s_3, \sigma_2} f_{{\bm k}_4, \ell,s_4,  \sigma_1}  \nonumber\\
&-& f_{{\bm    k}_{1}, \ell,s_1, \sigma_1} f_{{\bm k}_2, \ell,s_2,  \sigma_2} (1- f_{{\bm k}_3, \ell,s_3, \sigma_2}) (1-f_{{\bm k}_4, \ell, s_4, \sigma_1})]~.
\end{eqnarray}
\end{widetext}
The right-hand side of the previous equation is the collision integral and the Dirac delta distributions enforce conservation of momentum and energy in each e-e scattering event.
The quantity~\cite{Giuliani_and_Vignale}
\begin{equation}\label{eq:HF}
V^{(\ell,\sigma_{1},\sigma_{2})}_{1234} = V^{(\ell)}_{1234} - \delta_{\sigma_1, \sigma_2} V^{(\ell)}_{1243}
\end{equation}
in the collision integral includes a direct (Hartree) and an exchange (Fock) term, non-vanishing if two colliding electrons have parallel spins ($\sigma_{1} = \sigma_{2}$).
This expression for the kernel in the collision integral corresponds to the second-order Hartree-Fock approximation~\cite{KadanoffBaym}.
If spin-flip processes are absent (as in the case considered here), the distribution function does not depend on the spin label, which can be dropped.
The summation over $\sigma_{2}$ in Eq.~(\ref{eq:eescattering}) can be performed explicitly, obtaining the spin-independent kernel
\begin{eqnarray}\label{eq:matrixelementsummed}
& & |V_{s_{1},s_{2},s_{3},s_{4}}^{(\ell)}({\bm k}_{1}, {\bm k}_{2}, {\bm k}_{3}, {\bm k}_{4})|^{2} \nonumber \\
& & \equiv \sum_{\sigma_{2}} \left (1 - \frac{1}{2} \delta_{\sigma_{1},\sigma_{2}}  \right ) |V^{(\ell,\sigma_{1},\sigma_{2})}_{1234}|^{2} \nonumber \\
& & = \frac{1}{2}|V^{(\ell)}_{1234} - V^{(\ell)}_{1243}|^{2} + |V^{(\ell)}_{1234}|^{2}~,
\end{eqnarray}
in agreement with Ref.~\onlinecite{malic_prb_2011}.

\subsection{Electron-phonon interactions}

Electrons scatter with lattice vibrations and lose (gain) energy by emitting (absorbing) phonons.
Only optical phonons in the neighborhood of the $\Gamma$ and ${\rm K}$ points of the Brillouin zone (BZ) matter for electrons with energy of several hundred ${\rm meV}$ above the Fermi energy.
At each point, both the transverse (T) and the longitudinal (L) phonon modes are considered.
The distribution function of the $\nu$-th phonon mode with $\nu \in \{ \Gamma, {\rm L}; \Gamma, {\rm T}; {\rm K,L}; {\rm K,T} \}$ and 2d momentum ${\bm q}$ is denoted by the symbol $n_{\bm q}^{(\nu)}$.

The electron-phonon (e-ph) contribution to the EOM for the electron distribution is~\cite{MahanBook}
\begin{widetext}
\begin{eqnarray}\label{eq:electronphononboltzmann}
& & \left . \frac{d f_{{\bm k}, \ell, s}}{dt} \right |_{\rm e-ph} = -\frac{2 \pi}{\hbar} \frac{1}{A}~\sum_{\bm q}\sum_{s', \ell', \nu} {\cal A}_{0} |g^{(\nu)}_{{\bm k}, \ell \to {\bm k} + {\bm q},\ell'}|^{2}f_{{\bm k}, \ell, s}(1 - f_{{\bm k} + {\bm q}, \ell', s'})   \nonumber \\
& & \times \left[\delta (\varepsilon_{{\bm k} + {\bm q},s'} - \varepsilon_{{\bm k}, s} + \hbar \omega^{(\nu)}_{\bm q}) (n^{(\nu)}_{\bm q} + 1)
+ \delta (\varepsilon_{{\bm k} + {\bm q}, s'} - \varepsilon_{{\bm k}, s} - \hbar\omega^{(\nu)}_{\bm q})  n^{(\nu)}_{\bm q} \right]\nonumber \\
& & + \frac{2 \pi}{\hbar} \frac{1}{A}~\sum_{\bm q}\sum_{s',\ell', \nu} {\cal A}_{0} |g^{(\nu)}_{{\bm k} + {\bm q},\ell' \to {\bm k},\ell}|^{2} f_{{\bm k} + {\bm q},\ell', s'} (1 - f_{{\bm k}, \ell, s}) \nonumber \\
& & \times \left[\delta (\varepsilon_{{\bm k}, s} - \varepsilon_{{\bm k} + {\bm q}, s'} + \hbar\omega^{(\nu)}_{\bm q})  (n^{(\nu)}_{\bm q} + 1) + \delta (\varepsilon_{{\bm k}, s} - \varepsilon_{{\bm k} + {\bm q}, s'} - \hbar\omega^{(\nu)}_{\bm q}) n^{(\nu)}_{\bm q} \right]~,
\end{eqnarray}
\end{widetext}
where ${\cal A}_{0} \simeq 0.052~{\rm nm}^{2}$ is the area of the elementary cell of graphene's honeycomb lattice.
The terms proportional to $f_{{\bm k},\ell,s}(1-f_{{\bm k}',\ell',s'})$ represent electronic transitions from the single-particle state with quantum numbers ${\bm k}$, $\ell$, $s$, to the state ${\bm k}'$, $\ell'$, $s'$.
The transition is suppressed if the value $f_{{\bm k}',\ell',s'}$ of the distribution function in the final state is close to unity (Pauli blocking). The terms proportional to $n_{{\bm q}}^{(\nu)}$ correspond to absorption of phonons, while the terms proportional to $n_{{\bm q}}^{(\nu)} + 1$ correspond to emission of phonons.
The latter coefficient is larger than the former (Bose enhancement) because phonons, being bosonic excitations, experience bunching.
The kernels $g^{(\nu)}_{{\bm k},\ell \to {\bm k'},\ell'}$ can be written as:
\begin{eqnarray}\label{eq:elphcoupling}
|g^{(\Gamma, {\rm L})}_{{\bm k}, {\rm K} \to {\bm k} + {\bm q}, {\rm K}} |^{2}
& = & \langle g_{\Gamma}^{2} \rangle [1 + \cos{(\theta_{{\bm k}, {\bm q}} + \theta_{{\bm k} + {\bm q}, {\bm q}})}] \nonumber \\
|g^{(\Gamma, {\rm T})}_{{\bm k}, {\rm K} \to {\bm k} + {\bm q}, {\rm K}} |^{2}
& = & \langle g_{\Gamma}^{2} \rangle [1 - \cos{(\theta_{{\bm k}, {\bm q}} + \theta_{{\bm k} +{\bm q}, {\bm q}})}] \nonumber \\
|g^{({\rm K,L})}_{{\bm k}, {\rm K} \to {\bm k} + {\bm q}, {\rm K}'} |^{2}
& = & \langle g_{{\rm K},1}^{2} \rangle \nonumber \\
|g^{({\rm K,T})}_{{\bm k}, {\rm K} \to {\bm k} + {\bm q}, {\rm K}'} |^{2}
& = & \langle g_{{\rm K},2}^{2} \rangle [1 + \cos{(\theta_{{\bm k}, {\bm k} + {\bm q}})}]~,
\end{eqnarray}
where $\theta_{{\bm k},{\bm q}}$ denotes the angle between the wave vectors ${\bm k}$ and ${\bm q}$ and $\langle g_{X}^{2} \rangle$ are the electron-phonon couplings (EPCs)~\cite{PLMFR2004,BPF2009,lazzeri_prb_2008,ferrari_naturenanotech_2013}.
Phonons at the $\Gamma$ (${\rm K}$) point are responsible for intra-valley (inter-valley) scattering only.

The complete EOM for the electron distribution is the sum of Eqs.~(\ref{eq:eescattering}) and (\ref{eq:electronphononboltzmann}), i.e.:
\begin{equation}\label{eq:summingrates}
\frac{d f_{{\bm k}, \ell,s}}{dt} = \left . \frac{d f_{{\bm k},\ell,s}}{dt} \right |_{\rm e-e} + \left . \frac{d f_{{\bm k}, \ell, s}}{dt} \right |_{\rm e-ph}~.
\end{equation}
Finally, the SBE for the phonon distribution is:
\begin{eqnarray}\label{eq:phonon}
& & \frac{d n^{(\nu)}_{\bm q}}{dt}
= \frac{2 \pi}{\hbar}~\frac{1}{A} \sum_{\bm  k} \sum_{s,s',\ell,\ell'} {\cal A}_{0} \, |g^{(\nu)}_{{\bm k}, \ell \to {\bm k} + {\bm q}, \ell'}|^{2} \nonumber \\
& & \times f_{{\bm k}, \ell,s} (1 -f_{{\bm k} + {\bm q}, \ell', s'}) \left [ \vphantom{\sum} \delta (\varepsilon_{{\bm k} + {\bm q},  s'} - \varepsilon_{{\bm k}, s} + \hbar\omega^{(\nu)}_{\bm q}) \right .\nonumber \\
& & \left . \times (n^{(\nu)}_{\bm q} + 1) - \delta (\varepsilon_{{\bm k} + {\bm q},  s'} - \varepsilon_{{\bm k}, s} - \hbar\omega^{(\nu)}_{\bm q}) n^{(\nu)}_{\bm q} \vphantom{\sum} \right ] \nonumber \\
& & - \frac{\gamma_{\rm ph}}{\hbar} \left [ n^{(\nu)}_{\bm q} -  \frac{1}{\exp{(\hbar\omega^{(\nu)}_{\bm q} / (k_{\rm B} T_0))} - 1} \right ]~.
\end{eqnarray}
The right-hand side of the previous equation includes a phenomenological decay term which describes phonon-phonon interactions (due to the anharmonicity of the lattice).
Indeed anharmonic couplings play an important role in the graphene lattice~\cite{mounet_prb_2005,bonini_prl_2007, zakharchenko_prl_2009,zakharchenko_prb_2010,katsnelson_acr_2013} and, in principle, the decay coefficient $\gamma_{\rm ph}$ could be calculated by means of atomistic Monte Carlo simulations based on a realistic description of interatomic interactions~\cite{zakharchenko_prl_2009,zakharchenko_jphys_2011,katsnelson_acr_2013}.
The decay term induces relaxation of the phonon distribution towards the equilibrium value, given by a Bose-Einstein distribution at the temperature $T_0$ of the lattice.

\subsection{Relaxation timescales of a hot-electron distribution in graphene}
\label{ssec:timescales}

Accurate calculations of relaxation timescales in e.g.~semiconductors pose a challenging problem of great theoretical and practical relevance~\cite{HaugJauhoBook,snoke_annphys_2011}.
In graphene, three stages of the time evolution have been identified~\cite{sun_acsnano_2010,butscher_apl_2007,sun_prl_2008,dawlaty_apl_2008,breusing_prl_2009,obraztsov_nanolett_2011,brida_arxiv_2012},
which follow the creation of a HED due to the action of a laser-light ``pump'' pulse promoting a certain density of electrons from valence to conduction band.

In the first stage, $t \lesssim 20~{\rm fs}$, the initial HED thermalizes to a hot FD distribution and the two bands are characterized by different chemical potentials.
Recently, we were able to track this initial stage with sufficient time-resolution to directly measure the transition from a non-thermal to a hot FD distribution~\cite{brida_arxiv_2012}.
Cooling of the hot FD distribution and equilibration of the chemical potentials between the two bands take place in the second and third stage, where the dominant process is phonon emission.
The second stage, $t \lesssim 200~{\rm fs}$, is dominated by the emission of {\it optical} phonons~\cite{lazzeri_prl_2005}, which in graphene are associated with an unusually large energy scale ($\sim 200{\rm meV}$)~\cite{PLMFR2004,ferrari_prl_2006}
and are moderately coupled to the electronic degrees of freedom.
This cooling channel experiences a bottleneck when the phonon distribution heats up~\cite{lazzeri_prl_2005,butscher_apl_2007}.
The third stage, which occurs when the bulk of the electron distribution lies below the optical-phonon energy scale, is characterized by the emission of {\it acoustic} phonons~\cite{BM2009,tse_prb_2009,sun_acsnano_2010}.
These processes take place for $t \lesssim {\rm ns}$, but can experience a substantial speed-up ($t \sim 1~{\rm ns} \to 1~{\rm ps}$) in the case of disorder-assisted collisions~\cite{SRL2012,graham_naturephys_2012,betz_naturephys_2012}.
Since here we focus on the electron relaxation dynamics in the sub-$100~{\rm fs}$ time scale, we neglect the contribution of acoustic phonons in our SBE formulation.

Throughout the relaxation dynamics, phonons dissipate energy into the lattice by means of phonon-phonon interactions.
\section{Isotropic dynamics and collinear scattering processes}
\label{sect:collinearscattering}
\subsection{Semiclassical Boltzmann equation in the isotropic limit}
In this Section we simplify Eqs.~(\ref{eq:summingrates})-(\ref{eq:phonon}) by assuming that the electron and phonon distributions are {\it isotropic}.
While this assumption does not apply {\it during} the application of the pump pulse (since this couples anisotropically~\cite{malic_prb_2011}), it has been shown that e-e interactions restore isotropy in the very short time scale of a few fs~\cite{malic_prb_2011}.
Hence, our assumption applies to the thermalization and cooling stages of the time evolution, i.e. what we aim to study here.

The electron distribution $f_{\ell}(\varepsilon)$ is therefore assumed to depend on the wave vector ${\bm k}$ only through the energy $\varepsilon = \varepsilon_{{\bm k},s}$.
Similarly, the phonon distribution $n^{(\nu)}(q)$ is assumed to depend only on the magnitude $q$ of the phonon wave vector ${\bm q}$.
Since the slope of the phonon dispersion $\hbar \omega_{\bm q}^{(\nu)}$ is negligible with respect to $\hbar v_{\rm F}$, we drop the momentum dependence and use constant values $\omega^{(\Gamma)}$ and $\omega^{({\rm K})}$.
Equations for isotropic distributions can be obtained by performing the angular integrations in the collision integrals of Eqs.~(\ref{eq:summingrates}) and~(\ref{eq:phonon}). 

We now outline our approach in the case of a single summation over a wave vector ${\bm k}$, involving a generic function $g({\bm k},{\bm q})$, which depends on the direction of ${\bm k}$ and another wave vector ${\bm q}$, and a functional ${\cal F}[\varepsilon',\varepsilon'']$, which depends only on the isotropic quantities $\varepsilon' = \varepsilon_{{\bm k},s}$ and $\varepsilon'' = \varepsilon_{{\bm k} + {\bm q},s'}$.
We have~\cite{HaugJauhoBook}:
\begin{eqnarray}\label{eq:HaugTrick}
& & \sum_{\bm k} g({\bm k}, {\bm q}) {\cal F}[\varepsilon_{{\bm k},s},  \varepsilon_{{\bm k} + {\bm q},s'}] \nonumber \\
& & = \int_{-\infty}^{\infty} d\varepsilon' \, \int_{-\infty}^{\infty}d\varepsilon'' \, {\cal F}[\varepsilon',\varepsilon''] {\cal  Q}[\varepsilon',\varepsilon'']~,
\end{eqnarray}
where the kernel ${\cal Q}[\varepsilon',\varepsilon''] = \sum_{\bm k} \delta (\varepsilon' -\varepsilon_{{\bm k},s}) \delta (\varepsilon'' - \varepsilon_{{\bm k} +  {\bm q},s'}) g({\bm k}, {\bm q})$ depends now only on isotropic quantities.

The calculation of the isotropic kernels in Eqs.~(\ref{eq:summingrates}) and~(\ref{eq:phonon}) is summarized in the following.
This approach is convenient from a computational point of view since it reduces the number of variables in the integrations that have to be carried out numerically (see App.~\ref{ssec:numerical}).
Most importantly, it also allows us to handle analytically the contribution of collinear scattering to the e-e interaction in the Boltzmann collision integral.

The final results for the e-ph contributions are:
\begin{widetext}
\begin{eqnarray}\label{eq:phononsisotropic}
\frac{dn^{(\nu)}(q)}{dt} &=& \sum_{\ell,\ell'}\Bigg\{[n^{(\nu)}(q) + 1] \int_{-\infty}^{\infty} d\varepsilon f_{\ell}(\varepsilon)[1-f_{\ell'}(\varepsilon^{(\nu,-)})]{\cal Q}^{(\nu,-)}_{\ell,\ell'}(\varepsilon, q) \nonumber \\
&-& n^{(\nu)}(q) \int_{-\infty}^{\infty} d\varepsilon f_{\ell}(\varepsilon)[1-f_{\ell'}(\varepsilon^{(\nu,+)})]
{\cal Q}^{(\nu,+)}_{\ell,\ell'}(\varepsilon, q) \Bigg\} - \frac{\gamma_{\rm ph}}{\hbar} \left [ n^{(\nu)}(q) - \frac{1}{\exp{(\hbar\omega^{(\nu)} / (k_{\rm B} T_{0}))} - 1} \right ]
\end{eqnarray}
and
\begin{eqnarray}\label{eq:electronsphononsisotropic}
\left. \frac{df_{\ell}(\varepsilon)}{dt} \right|_{\rm e-ph} &=& \frac{2\pi}{\hbar} \sum_{\ell',\nu}
\Bigg\{
- f_{\ell}(\varepsilon) [1-f_{\ell'}(\varepsilon^{(\nu,-)})] \int_{0}^{\infty} dq (\hbar v_{\rm F})[n^{(\nu)}(q) + 1]~{\cal  I}^{(\nu,-)}_{\ell,\ell'}(\varepsilon,q) \nonumber\\
&-& f_{\ell}(\varepsilon)[1-f_{\ell'}(\varepsilon^{(\nu,+)})]\int_{0}^{\infty} dq (\hbar v_{\rm F}) n^{(\nu)}(q)~{\cal  I}^{(\nu,+)}_{\ell,\ell'}(\varepsilon,q) \nonumber \\
&+& f_{\ell'}(\varepsilon^{(\nu,+)})  [1-f_{\ell}(\varepsilon)]\int_{0}^{\infty} dq (\hbar v_{\rm F}) [n^{(\nu)}(q) + 1]~{\cal  I}^{(\nu,+)}_{\ell,\ell'}(\varepsilon,q) \nonumber \\
&+& f_{\ell'}(\varepsilon^{(\nu,-)}) [1-f_{\ell}(\varepsilon)] \int_{0}^{\infty} dq (\hbar v_{\rm F}) n^{(\nu)}(q) ~ {\cal  I}^{(\nu,-)}_{\ell,\ell'}(\varepsilon,q)\Bigg\}~,
\end{eqnarray}
\end{widetext}
with the shorthand $\varepsilon^{(\nu,\pm)} = \varepsilon \pm \hbar \omega^{(\nu)}$.

In Eqs.~(\ref{eq:phononsisotropic})-(\ref{eq:electronsphononsisotropic}), we introduced the following functions:
\begin{equation}\label{eq:PhononKernels}
{\cal I}^{(\nu,\pm)}_{\ell,\ell'}(\varepsilon, q)
= \frac{1}{\pi\hbar} \sum_{s'}I \left[ s' \frac{\varepsilon^{(\nu,\pm)}}{\hbar v_{\rm F} q}, \frac{|\varepsilon |}{\hbar v_{\rm F} q}, \frac{{\cal A}_{0} |{g}^{(\nu)}_{\ell,\ell'}|^{2}}{\hbar v_{\rm F}} \right ]
\end{equation}
and
\begin{equation}
{\cal Q}^{(\nu,\pm)}_{\ell,\ell'}(\varepsilon, q) = \frac{|\varepsilon|}{\hbar v_{\rm F} q} {\cal I}^{(\nu,\pm)}_{\ell,\ell'}(\varepsilon, q)~,
\end{equation}
with
\begin{eqnarray}
& & I[x_{0},x_{1},{\cal F}] \equiv \Theta(x_{0} - |x_{1} - 1|) \Theta(x_{1} + 1 - x_{0}) \nonumber \\
& & \times \frac{2 x_{0}}{\sqrt{4x_{1}^{2}-(x_{0}^{2}-x_{1}^{2}-1)^{2}}} \nonumber \\
& & \times {\cal F} \left ( x_{1},\arccos[(x_{0}^{2} - x_{1}^{2} - 1)/(2 x_{1})] \right )~.
\end{eqnarray}
Here $\Theta(x)$ is the Heaviside distribution and the quantities $x_{0}, x_{1}$ are dimensionless.
For notational convenience, we write the wave vector dependence of the e-ph kernel $|g^{(\nu)}_{{\bm k},\ell \rightarrow{\bm k} + {\bm q},\ell'}|^{2}$ in the form $|g^{(\nu)}_{\ell,\ell'}|^{2}(r,\theta)$, where $r = k / q$ and $\theta = \theta_{{\bm k},{\bm q}}$.

Finally, the e-e contribution reads
\begin{eqnarray}\label{eq:finalboltzmann}
& & \left . \frac{d f_{\ell}(\varepsilon_{1})}{dt} \right |_{\rm e-e}
= \int_{-\infty}^{+\infty}d\varepsilon_{2} \int_{-\infty}^{+\infty} d\varepsilon_{3} ~ {\cal C}^{(\ell)}(\varepsilon_{1}, \varepsilon_{3}, E) \nonumber \\
& & \times \left \lbrace [1 - f_{\ell}(\varepsilon_{1})] [1 - f_{\ell}(\varepsilon_{2})]f_{\ell}(\varepsilon_{3}) f_{\ell}(\varepsilon_{4}) \right . \nonumber \\
& & -  \left . f_{\ell}(\varepsilon_{1}) f_{\ell}(\varepsilon_{2}) [1 -  f_{\ell}(\varepsilon_{3})][1 - f_{\ell}(\varepsilon_{4})] \right \rbrace~.
\end{eqnarray}
where the Coulomb kernel ${\cal C}^{(\ell)}$, with physical dimensions ${\rm fs}^{-1}~{\rm eV}^{-2}$, represents a two-particle scattering rate.
The energies of the incoming (with indexes $1$ and $2$) and outgoing particles (with indexes $3$, $4$) are fixed.
The total energy $E \equiv \varepsilon_{1} + \varepsilon_{2}$ is conserved and, finally, $\varepsilon_{4} \equiv E - \varepsilon_{3}$.
\subsection{The Coulomb kernel}
\label{ssec:coulombkernel}

\begin{figure}
\includegraphics[width=\columnwidth]{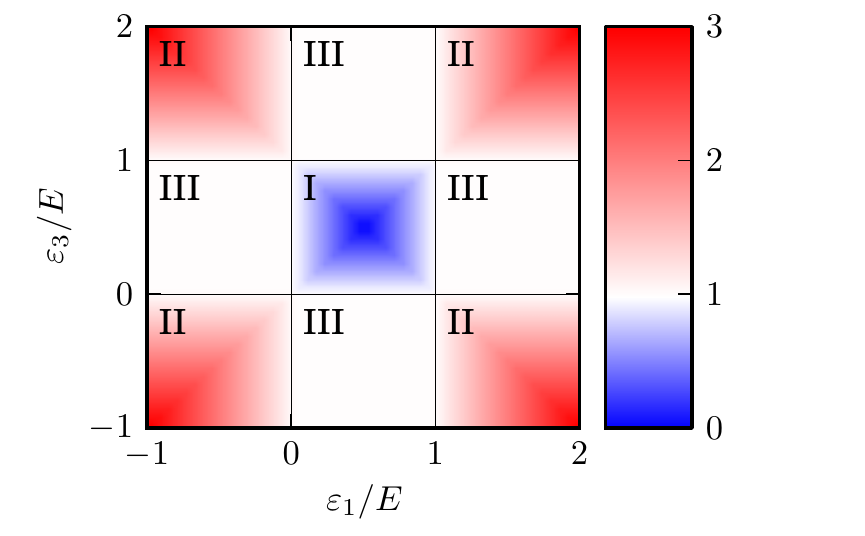}
\caption{(Color online) The integration domain~(Eq.\ref{eq:integrationdomain}) for the variable $Q$ in the integral~(Eq.\ref{eq:kernelintegralsimplified}).
The regions labeled by I, II, and III identify the values of the parameters $\varepsilon_1$ and $\varepsilon_3$ for which intra-band, inter-band, and Auger processes take place, respectively.
At $Q = |E| / (\hbar v_{\rm F})$ collinear scattering takes place.
This value is the maximum (minimum) of the integration domain in region I (II).
The minimum (maximum) of the integration domain in region I (II) is shown in the color scale, in units of $|E| / (\hbar v_{\rm F})$. \label{fig:IntegrationDomain}}
\end{figure}

Simplifying Eq.~(\ref{eq:eescattering}) along the lines of Eq.~(\ref{eq:HaugTrick}) leads to the following expression for the Coulomb kernel:
\begin{widetext}
\begin{eqnarray}\label{eq:coulombkernel}
{\cal C}^{(\ell)}(\varepsilon_{1}, \varepsilon_{3}, E)
&\equiv& \frac{2\pi}{\hbar} \lim_{\eta \to 0} \frac{1}{A^2} \sum_{{\bm Q}, {\bm k}_{3}}
\delta(|E - \varepsilon_{1}| - \hbar v_{\rm F} |{\bm Q} - {\bm k}_{1}|)~\delta(|\varepsilon_{3}| - \hbar v_{\rm F} k_{3})~\delta(|E - \varepsilon_{3}| - \hbar v_{\rm F} |{\bm Q} - {\bm k}_{3}| + \eta) \nonumber \\
&\times &  |V_{s_1, s_2, s_3, s_4}^{(\ell)}({\bm k}_{1}, {\bm Q} - {\bm k}_{1}, {\bm  k}_{3}, {\bm Q} - {\bm k}_{3})|^2 \vphantom{\sum}~.
\end{eqnarray}
\end{widetext}
Here, the wave vector ${\bm k}_{1}$ has modulus $\varepsilon_{1} / (\hbar v_{\rm F})$, while its direction can be fixed at will (i.e.~along the $\hat{\bm x}$ axis) because the final result ${\cal C}^{(\ell)}(\varepsilon_{1}, \varepsilon_{3}, E)$ is scalar under rotations.
The total wave vector ${\bm Q} = {\bm k}_{1} + {\bm k}_{2}$ is conserved in all scattering processes.
In the summation over ${\bm Q}$ and ${\bm k}_{3}$, the Dirac delta distributions ensure that only scattering configurations that are compatible with the choice of incoming $\varepsilon_{1}$ and outgoing $\varepsilon_{3}$ energies are considered.
Moreover, the three delta distributions restrict the 4d integral to a 1d integral (at most), after the usual continuum limit $A^{-1} \sum_{\bm k} \to (2\pi)^{-1} \int d {\bm k}^{2}$ is performed.
We choose to reduce the summations to an integration over the modulus $Q$ of the total momentum, in terms of which we are able to represent with clarity the phase space available for Coulomb scattering (see Fig.~\ref{fig:IntegrationDomain}).

We stress that in Eq.~(\ref{eq:coulombkernel}) we introduced an {\it infinitesimal} quantity $\eta$ in the argument of one of the delta distributions.
As we will see in the next Section, if the limit $\eta \to 0$ is taken {\it before} calculating the 4d integral in Eq.~(\ref{eq:coulombkernel}), collinear scattering processes do not contribute to ${\cal C}^{(\ell)}(\varepsilon_{1}, \varepsilon_{3}, E)$.
We introduced $\eta$ to slightly relax the condition of energy conservation.
The latter is recovered only in the limit $\eta \to 0$.
We can justify this by considering the following physical explanation.
The delta distribution of conservation of energy in Eq.~(\ref{eq:eescattering}) originates from the so-called ``quasiparticle approximation,'' applied to the Kadanoff-Baym equations (KBEs), from which the SBE is derived~\cite{KadanoffBaym}.
More precisely, the KBEs involve the true quasiparticle spectral function, which has a finite width.
In the quasiparticle approximation, the spectral function is substituted with a delta distribution, which is a reasonable approximation when the width of the quasiparticle spectral function can be neglected.
As we will see in this Section, in the quasiparticle approximation applied to 2d MDFs an entire class of two-body collisions (collinear processes) yields vanishing scattering rates.
Our procedure takes effectively into account the fact that quasiparticles have a finite lifetime~\cite{bostwick_naturephys_2007, zhou_naturemater_2007,bostwick_science_2010,walter_prb_2011,siegel_pnas_2011, knox_prb_2011}, thereby allowing for a finite collinear scattering contribution to the Coulomb kernel.
We first calculate the Coulomb kernel with a finite $\eta$ and then apply the quasiparticle $\eta \to 0$ approximation at the end of the calculation.

To make analytical progress, we now introduce elliptic coordinates for the evaluation of the Coulomb kernel~\cite{Sachdev1998}.
This is most natural because, for every fixed value of ${\bm Q}$, the equation $E = s_{1} \hbar v_{\rm F} k_{1} + s_{2} \hbar v_{\rm F} k_{2}$ for the total energy defines a conic section in momentum space.
More precisely, if $s_{1} = s_{2}$ ($s_{1} \neq s_{2}$), the vector ${\bm k} = {\bm k}_{1} - {\bm Q}/2$ lies on an ellipse (hyperbola) with focuses located at $\pm {\bm Q} / 2$ and major axis of length $|E| / (\hbar v_{\rm F})$.
Elliptic coordinates $(u,v)$ are related to the Cartesian coordinates $(k_{x},k_{y})$ by the transformation $k_{x} = (Q / 2) \cosh{(u)} \cos{(v)}$, $k_{y} = (Q / 2) \sinh{(u)} \sin{(v)}$, with area element $d{\bm k}^{2} = (Q/2)^{2}[ \sinh{(u)}^{2} + \sin{(v)}^{2} ] \, du \, dv$.
In these coordinates, $k_{i} = (Q/2)[\cosh{(u_{i})} + \cos{(v_{i})}]$ and $|{\bm Q} - {\bm k}_{i}| = (Q/2)[\cosh{(u_{i})} - \cos{(v_{i})}]$, so that non-linear combinations between integration variables in Eq.~(\ref{eq:coulombkernel}) disappear.
Elliptic coordinates are also extremely useful to prove that IMI and AR can only occur when ${\bm k}_{1},\dots, {\bm k}_{4}$ are collinear (see Figs.~4a and~b in Ref.~\onlinecite{brida_arxiv_2012}).

Carrying out algebraic manipulations, we rewrite Eq.~(\ref{eq:coulombkernel}) in the following simplified manner:
\begin{widetext}
\begin{eqnarray}\label{eq:kernelintegralsimplified}
& & {\cal C}^{(\ell)}(\varepsilon_{1},\varepsilon_{3},E) = \frac{2 \pi}{\hbar} \frac{1}{(2\pi)^{4}} \frac{1}{(\hbar v_{\rm F})^{3}}\lim_{\eta \to 0}
\int_{Q_{0}}^{Q_{1}} dQ \sideset{}{'}\sum |V^{(\ell)}_{s_{1},s_{2},s_{3},s_{4}}({\bm k}_{1},{\bm k}_{2},{\bm k}_{3}, {\bm k}_{4})|^{2} \nonumber \\
& &
\times \frac{(E_{1,{\rm max}} - E_{1,{\rm min}})(\hbar v_{\rm F} Q)}{\left \lbrace [E_{1,{\rm max}}^{2} - (\hbar v_{\rm F} Q)^{2}][(\hbar v_{\rm F} Q)^{2} - E_{1,{\rm min}}^{2} ] \right \rbrace ^{1/2}}
\frac{ (E_{3,{\rm max}} + \eta)^{2} - (E_{3,{\rm min}} - \eta)^{2}
}{\left \lbrace [(E_{3,{\rm max}}+\eta)^{2} - (\hbar v_{\rm F} Q)^{2}][(\hbar v_{\rm F} Q)^{2} - (E_{3,{\rm min}} - \eta)^{2} ] \right \rbrace ^{1/2}}~. \nonumber \\
\end{eqnarray}
\end{widetext}
We stress that this is the most important analytical result of this Article.

Note that the integrand in Eq.~(\ref{eq:kernelintegralsimplified}) is given by the product of the kernel (\ref{eq:matrixelementsummed}) and a complicated expression arising from the phase space of the e-e scattering processes, the term in the second line of Eq.~(\ref{eq:kernelintegralsimplified}).
In Eq.~(\ref{eq:kernelintegralsimplified}), $s_{i} = {\rm sgn}(\varepsilon_{i})$ and the dependence of ${\bm k}_{i}$ on $\varepsilon_{1}$, $\varepsilon_{3}$, $E$, and $Q$ is left implicit for the sake of simplicity.
The primed sum symbol indicates summation over the available configurations of vectors ${\bm k}_{2}$, ${\bm k}_{3}$, and ${\bm k}_{4}$. To identify these configuration, one may proceed as follows.
When $\varepsilon_{1}$, $\varepsilon_{3}$, $E$, and $Q$ are given, the lengths of all the sides of the two triangles $({\bm k}_{1}, {\bm k}_{2}, {\bm Q})$ and $({\bm k}_{3}, {\bm k}_{4}, {\bm Q})$ are uniquely fixed.
These two triangles, which share the side of length $Q$, can be drawn on the same half-plane (with respect to a line containing the vector ${\bm Q}$) or on opposite half-planes.
Thus, four geometric configurations for the wave vectors ${\bm k}_{2}$, ${\bm k}_{3}$, and ${\bm k}_{4}$ are available in total.
However, when all vectors ${\bm k}_{i}$ are collinear, the triangles are degenerate and only one configuration is possible.
Finally, in Eq.~(\ref{eq:kernelintegralsimplified}) we also introduced $E_{i, {\rm min}} = |\varepsilon_{i}| - |E - \varepsilon_{i}|$, $E_{i, {\rm max}} =  |E - \varepsilon_{i}| + |\varepsilon_{i}|$, and
\begin{eqnarray}\label{eq:integrationdomain}
\hbar v Q_{0} & = & {\rm max} \left ( |E_{1,{\rm min}}|, |E_{3,{\rm min}} - \eta| \right ) \nonumber \\
\hbar v Q_{1} & = & {\rm min} \left ( E_{1,{\rm max}}, E_{3,{\rm max}} + \eta \right ) ~.
\end{eqnarray}
Let us first discuss the case in which $\eta$ is set to zero before carrying out the integral in Eq.~(\ref{eq:kernelintegralsimplified}).
In this case, one can prove that $Q_{0} \le Q_{1}$ by using the triangular and reverse triangular inequalities, $ ||E - \varepsilon_{i}| - |\varepsilon_{i}|| \le |E| \le |E - \varepsilon_{j}| + |\varepsilon_{j}|$.
When the previous inequalities turn into equalities, the length of the vector ${\bm Q}$ is fixed at $Q = |E| / (\hbar v_{\rm F})$, $Q_{0} = Q_{1}$, and the integration domain vanishes.
Fig.~\ref{fig:IntegrationDomain} plots the integration domain relative to the variable $Q$ in Eq.~(\ref{eq:integrationdomain}), as a function of $\varepsilon_{1}$ and $\varepsilon_{3}$, in the case $\eta = 0$.
In region I, $0 < \varepsilon_{1}, \varepsilon_{3} < E$ implies $0 < \varepsilon_{2}, \varepsilon_{4} < E$ (if the total energy is negative, all the inequalities are reversed).
All the particles are either above ($E > 0$) or below ($E < 0$) the Dirac point.
Region I, therefore, pertains to intra-band scattering events.
Similarly, one concludes that regions of type II pertain to inter-band scattering (two electrons are in opposite bands before and after the scattering).
Finally, regions of type III pertain to IMI and AR.
In these regions the integration domain [and, therefore, ${\cal C}^{(\ell)}(\varepsilon_{1},\varepsilon_{3},E)$] in Eq.~(\ref{eq:kernelintegralsimplified}) vanishes.
We note that classification of regions I, II, and III holds true for arbitrary values of $\eta$.

Eq.~(\ref{eq:kernelintegralsimplified}) is extremely helpful since it can be used to solve the SBE~(\ref{eq:summingrates}) with arbitrary non-equilibrium initial conditions, more so since analytical expressions for the Coulomb kernel of 2d MDFs such as that in Eq.~(\ref{eq:kernelintegralsimplified}) were not reported before, to the best of our knowledge.

\subsection{Auger contribution to the Coulomb collision integral}
\label{sect:Auger}

We now proceed to calculate ${\cal C}^{(\ell)}(\varepsilon_{1},\varepsilon_{3},E)$ in regions of type III.
In this case, a finite value of $\eta$ restores a non-vanishing integration domain for IMI and AR and a finite contribution to ${\cal C}^{(\ell)}(\varepsilon_{1},\varepsilon_{3},E)$ due to these processes.
Note that the sign of $\eta$ should be chosen such that $Q_{0} < Q_{1}$.
Let us consider, for the sake of definiteness, the region of type III where $0 < \varepsilon_{1} < E < \varepsilon_{3}$.
We have $Q_{0} = (E - |\eta|) / (\hbar v_{\rm F}) < E / (\hbar v_{\rm F}) = Q_{1}$.
The integrand in Eq.~(\ref{eq:kernelintegralsimplified}) factors into two portions, one that depends smoothly on $Q$ and can therefore be evaluated at $Q = E / (\hbar v_{\rm F})$ and taken out of the integral, and another singular at the boundaries of the integration domain.
The integral of the latter part must be carefully evaluated and is:
\begin{equation}\label{eq:abeautifulintegral}
\int_{Q_0}^{Q_1}
\frac{d Q}{[(Q_1 - Q) (Q - Q_0)]^{1/2}} = \pi~.
\end{equation}
Note that the result of the previous integral does {\it not} depend on $\eta$, therefore remains finite in the limit $\eta \to 0$.

The final result for the Auger contribution to the Coulomb kernel, valid in {\it all} regions of type III, can be written as:
\begin{eqnarray}\label{eq:AugerResult}
& & \left.{\cal C}^{(\ell)}(\varepsilon_{1},\varepsilon_{3},E)\right|_{\rm Auger} = \frac{1}{8\pi^2 \hbar^5 v_{\rm F}^4} \sqrt{ \left | \frac{\varepsilon_{2}\varepsilon_{3}\varepsilon_{4}}{\varepsilon_{1}} \right |} \nonumber \\
& & \times |V^{(\ell)}_{s_{1},s_{2},s_{3},s_{4}}({\bm k}_{1},{\bm k}_{2},{\bm k}_{3}, {\bm k}_{4})|^{2}~,
\end{eqnarray}
where the convention for $s_{i}$ and ${\bm k}_{i}$ has been introduced after Eq.~(\ref{eq:kernelintegralsimplified}).
The term on the second line of Eq.~(\ref{eq:AugerResult}) comes from the smooth portion of the integrand in Eq.~(\ref{eq:kernelintegralsimplified}).
We stress that Eq.~(\ref{eq:AugerResult}) follows from the general expression~(\ref{eq:kernelintegralsimplified}) without \emph{a priori} restrictions to collinear scattering configurations.
Although Eq.~(\ref{eq:AugerResult}) mathematically coincides with Eq.~(14) of Ref.~\onlinecite{rana_prb_2007}, Ref.~\onlinecite{rana_prb_2007} does not report any discussion on how to bypass the vanishing phase space problem for 2d MDFs.
Here, the finiteness of IMI and AR contributions to the Coulomb kernel, as for Eq.~(\ref{eq:AugerResult}), originates from electron-lifetime effects.
Incidentally, since the value of the integral in Eq.~(\ref{eq:abeautifulintegral}) does not depend on the value of $\eta$, the precise mechanism (e-e interactions, electron-impurity scattering, etc.) responsible for the broadening of the delta distribution in Eq.~(\ref{eq:eescattering}) into a finite-width quasiparticle spectral function is unimportant.
Finally, we emphasize that IMI and AR scattering rates were calculated in Ref.~\onlinecite{rana_prb_2007} for FD distributions only, as seen in Eq.~(20) of Ref.~\onlinecite{rana_prb_2007}.
On the contrary, Eqs.~(\ref{eq:kernelintegralsimplified}) and~(\ref{eq:AugerResult}) can be used to solve the SBE~(\ref{eq:summingrates}) with {\it arbitrary} non-equilibrium initial conditions.

\subsection{Logarithmically divergent collinear scattering rates}
\label{ssec:divergencies}

We finally consider ${\cal C}^{(\ell)}(\varepsilon_{1},\varepsilon_{3},E)$ in regions of type I and II.
The integrand in Eq.~(\ref{eq:kernelintegralsimplified}) diverges as $|\hbar v_{\rm F} Q - |E||^{-1}$ for $Q = |E|/(\hbar v_{\rm F})$, which coincides with the upper $Q_1$ or lower $Q_0$ boundaries of the integration domain for regions of type I  and II.
When $Q = |E|/(\hbar v_{\rm F})$, intra-band or inter-band scattering occur in a collinear fashion.
This strong divergence of the integrand physically arises from the expression for the phase space of e-e scattering, while $V^{(\ell)}_{s_{1},s_{2},s_{3},s_{4}}({\bm k}_{1},{\bm k}_{2},{\bm k}_{3}, {\bm k}_{4})|^{2}$ is well behaved.
Therefore, ${\cal C}^{(\ell)}(\varepsilon_{1},\varepsilon_{3},E)$ diverges for both intra-band and inter-band scattering processes.
A possible way to cure this divergence~\cite{Sachdev1998,kashuba_prb_2008,fritz_prb_2008} is to introduce an ultraviolet cut-off $\Lambda$, which yields a Coulomb kernel $\propto \ln(\Lambda)$. Logarithmic enhancements for 2d Fermions with a linear dispersion were discussed in Ref.~\onlinecite{Sachdev1998}, and allow one to find a SBE solution in the form of an effective equilibrium distribution, with parameters depending on the direction of motion~\cite{fritz_prb_2008}.
Peculiar properties of MDFs, which are sensitive to collinear scattering, include a finite conductivity in the absence of impurities~\cite{kashuba_prb_2008} and an unusually low shear viscosity~\cite{MSF2009}.

A different way to treat this divergence is to invoke {\it screening}, which suppresses the kernel $|V^{(\ell)}_{s_{1},s_{2},s_{3},s_{4}}({\bm k}_{1},{\bm k}_{2},{\bm k}_{3}, {\bm k}_{4})|^{2}$ and regularizes the behavior of the integrand in a neighborhood of $Q = |E| / (\hbar v_{\rm F})$. This approach is discussed in the next Section in great detail.

\section{Going beyond the Fermi golden rule: the role of screening}
\label{sect:screening}

The SBE is a second-order expression in the {\it bare} Coulomb potential $v_{q}$ and describes e-e interactions at level of the Fermi golden rule~\cite{Giuliani_and_Vignale}.
This approximation neglects many-body effects, and most importantly electronic screening.
Formally, screening can be taken into account~\cite{Giuliani_and_Vignale} by substituting the bare Coulomb potential $v_q$ with a {\it screened} potential $W$.
When the 2d MDF system is out of equilibrium, the screening properties change in time and the screened potential depends on time $t$ as well.

It has been pointed out~\cite{brida_arxiv_2012,SOTGNM2013,STKL2012} that screening may preempt the strong collinear scattering singularity mentioned above and suppress Auger processes.
Indeed, the RPA dynamical dielectric function at equilibrium $\epsilon(q,\omega)$ diverges for collinear scattering configurations for which $\omega = \pm v_{\rm F} q$ (see Sec.~\ref{ssec:dielectric}).
When substituted into Eq.~(\ref{eq:kernelintegralsimplified}), the screened potential $W$ vanishes like $|\hbar v_{\rm F} Q - |E||^{1/2}$, thereby compensating the aforementioned divergence arising from the expression of the phase space.
The integral in Eq.~(\ref{eq:kernelintegralsimplified}) is then finite, while the contribution to the Coulomb kernel due to Auger processes vanishes.
\subsection{Time-dependent dielectric screening in a photo-excited 2d MDF fluid}
\label{ssec:dielectric}

The matrix element of the screened potential is obtained by the replacement:
\begin{equation}\label{eq:screenedpotential}
V^{(\ell)}_{1234} \to W^{(\ell)}_{1234}  = \left.\frac{V^{(\ell)}_{1234}}{\epsilon(q,\omega)}\right|_{
\begin{array}{l}
q = |{\bm k}_{1} - {\bm k}_{3}|\\
\omega = (\varepsilon_1 -\varepsilon_3)/\hbar
\end{array}}~,
\end{equation}
where $V^{(\ell)}_{1234}$ is defined in Eq.~(\ref{eq:matrixinteraction}) and $\epsilon(q,\omega)$ is the dynamical dielectric function~\cite{Giuliani_and_Vignale}.
Here, $\hbar q = \hbar|{\bm k}_{1} - {\bm k}_{3}|$ and $\hbar \omega = \varepsilon_1 -\varepsilon_3$ are the momentum and energy transferred in the scattering process, respectively.

We stress that the prescription~(\ref{eq:screenedpotential}) must be applied only to the first term on the right hand side of Eq.~(\ref{eq:HF}), i.e.~to the {\it direct} term.
In principle, one could apply Eq.~(\ref{eq:screenedpotential}) to screen both direct and exchange contributions in Eq.~(\ref{eq:HF}).
The latter procedure was previously used in Refs.~\onlinecite{Collet1993,HAB1999,LG2000} to compute scattering rates, but has a major drawback: it is easy to see that the corresponding SBE does {\it not} conserve the particle number.
In other words, the approximation obtained by screening both direct and exchange terms in Eq.~(\ref{eq:HF}) according to Eq.~(\ref{eq:screenedpotential}) is not conserving in the sense of Kadanoff and Baym~\cite{BK1961,KadanoffBaym}.
On the contrary, retaining the direct term only corresponds to the well-known ``shielded potential approximation"~\cite{KadanoffBaym}, which is conserving in the sense of Kadanoff and Baym.
This is the approach we follow below, setting $V_{1243}^{(\ell)} = 0$ in Eq.~(\ref{eq:HF}).

In the RPA, the dielectric function is given by~\cite{Giuliani_and_Vignale}
\begin{equation}
\epsilon(q,\omega) = 1 - v_{q} \chi^{(0)}(q,\omega)~,
\end{equation}
where $v_{q}$ is the bare Coulomb potential, defined in Eq.~(\ref{eq:Coulomb}), and $\chi^{(0)}(q,\omega)$ is the non-interacting polarization function~\cite{Giuliani_and_Vignale} (or Lindhard function) for 2d MDFs:
\begin{eqnarray}\label{eq:polarization}
& & \chi^{(0)}(q,\omega) = N_{\rm s} \sum_{\ell} \sum_{s s'} \int \frac{d^{2} {\bm k}}{(2 \pi)^{2}} M_{s,s'}({\bm k},{\bm k} + {\bm q}) \nonumber \\
& &
\times \frac{
f_{\ell}(\varepsilon_{{\bm k}, s}) - f_{\ell}(\varepsilon_{{\bm k} +{\bm q}, s'})
}{\hbar \omega + \varepsilon_{{\bm k}, s} - \varepsilon_{{\bm k} + {\bm q}, s'} + i \eta}~,
\end{eqnarray}
where $M_{s,s'}({\bm k},{\bm k} + {\bm q}) = [1 + s s' \cos{(\theta_{{\bm k} + {\bm q}} - \theta_{{\bm k}})}]/2$ and the factor $N_{\rm s} =2$ accounts for spin degeneracy.
It is intended that $\epsilon(q,\omega)$ and $\chi^{(0)}(q,\omega)$ depend {\it explicitly on time} $t$ through the time-dependence of the distribution function $f_{\ell}(\varepsilon)$.

One route to include screening in the SBE calculations is to compute~\cite{sun_prb_2012} the polarization function $\chi^{(0)}(q,\omega)$ at each time $t$ according to Eq.~(\ref{eq:polarization}) and use it to evaluate the expressions (\ref{eq:kernelintegralsimplified}) and (\ref{eq:AugerResult}) for the Coulomb kernel.
In this Article, however, we prefer to use a more analytical approach, which turns out to reduce dramatically the computational costs associated with solving the SBE with screening.
Since thermalization occurs on a very fast time scale~\cite{brida_arxiv_2012}, we calculated the polarization function analytically by employing the following {\it thermal} Ansatz for the distribution function $f_{\ell}(\varepsilon)$ in Eq.~(\ref{eq:polarization}):
\begin{eqnarray}\label{eq:distroTwoMu}
f_{\ell}(\varepsilon) &\to& \Theta(\varepsilon) F(\varepsilon; \mu_{\rm c}, T) + \Theta(-\varepsilon) F(\varepsilon; \mu_{\rm v}, T)\nonumber \\
&\equiv & F(\varepsilon; \mu_{\rm c}, \mu_{\rm v}, T)~,
\end{eqnarray}
where
\begin{equation}\label{eq:FDD}
F(\varepsilon;\mu,T) = \frac{1}{\exp{[(\varepsilon - \mu) / (k_{\rm B} T)]} + 1}
\end{equation}
is the usual FD distribution.
In Eq.~(\ref{eq:distroTwoMu}), $\mu_{\rm c}$ and $\mu_{\rm v}$ are chemical potentials in conduction and valence band, respectively, and $T$ is the temperature.

The values of the three parameters $\mu_{\rm c}$, $\mu_{\rm v}$, and $T$ can be obtained at each time $t$ by fitting Eq.~(\ref{eq:distroTwoMu}) to the HED derived by solving the SBE.
In writing Eq.~(\ref{eq:distroTwoMu}) we assumed that conduction- and valence-band electrons thermalize at the same $T$.
This approximation is certainly valid for times longer than $\simeq 20~{\rm fs}$ (see Sec.~\ref{ssec:timescales}), because energy equilibration between the two bands sets in on the time scale induced by e-e interactions.
At earlier times, the estimate of $T$ and $\mu_{s}$ obtained by fitting the profile (\ref{eq:distroTwoMu}) to the HED is certainly not precise, but can be improved {\it a posteriori} by extrapolating to $t \lesssim 20~{\rm fs}$ the results of the fits obtained at later times.
Although a common $T$ between the two bands is established in the very early stages of the dynamics, carrier equilibration between the two bands, on the contrary, is mainly due to phonon-assisted inter-band transitions, which act on a much longer time scale ($\gtrsim 200~{\rm fs}$).
Assuming two different chemical potentials is thus essential to obtain a correct representation of screening in the non-equilibrium dynamics after photo-excitation.
Our analytical approach to the calculation of $\chi^{(0)}(q,\omega)$ allows us to describe well the rapidly changing dielectric function and, at the same time, to determine analytically its behavior for collinear configurations, crucial to the issues discussed in Sec.~\ref{ssec:divergencies}.

In the rest of this Section we outline the calculation of the polarization function of a photo-excited 2d MDF fluid.
We start by noting that the polarization function $\chi^{(0)}(q,\omega)$ as obtained from Eq.~(\ref{eq:polarization}) with the thermal Ansatz~(Eq.\ref{eq:distroTwoMu}) physically represents the polarization function $\chi^{(0)}(q,\omega; T)$ of a 2d MDF fluid with {\it two} chemical potentials $\mu_{\rm c}$ and $\mu_{\rm v}$, at a finite $T$.
To the best of our knowledge, this function is unknown.

We therefore proceed to calculate $\chi^{(0)}(q,\omega; T)$ following a well-known procedure due to Maldague~\cite{Giuliani_and_Vignale,maldague_ss_1978}.
This route allows us to write this polarization function in terms of an integral involving the well-known Lindhard function $\Pi_0(q,\omega)$ of a 2d MDF fluid at zero $T$ and at a Fermi energy $\varepsilon_{\rm F}$~\cite{wunsch_njp_2006, hwang_prb_2007_lindhard,barlas_prl_2007,principi_prb_2009}.
The final result is:
\begin{widetext}
\begin{eqnarray}\label{eq:lindhardTwoMu}
\chi^{(0)}(q,\omega; T) &=& \int_{-\infty}^{\infty}
d\varepsilon' \left. \Pi_0(q, \omega)\right|_{\varepsilon_{\rm F} \to  \varepsilon'}\left [\Theta(\varepsilon'){\cal G}(\varepsilon';\mu_{\rm c}, T) + \Theta(-\varepsilon'){\cal G}(\varepsilon';\mu_{\rm v},T)\right ] \nonumber \\
& & - \left. \Pi_0(q,\omega)\right|_{\varepsilon_{\rm F} \to 0} \left [ F(0;\mu_{\rm c},T) - F(0;\mu_{\rm v},t) \right ]~,
\end{eqnarray}
\end{widetext}
where
\begin{equation}\label{eq:inputlindhardTwoMu}
{\cal G}(\varepsilon';\mu_s,T) =  \frac{1}{4 k_{\rm B} T \cosh^{2}{\left (\displaystyle \frac{\varepsilon' - \mu_s}{2 k_{\rm B} T} \right )}}~.
\end{equation}
Note that $\Pi_0(q,\omega)$ is particle-hole symmetric (therefore identical for positive and negative values of the Fermi energy $\varepsilon_{\rm F}$) and that $\left.\Pi_0(q,\omega)\right|_{\varepsilon_{\rm F} \to 0}$ is the Lindhard function of an undoped 2d MDF system~\cite{wunsch_njp_2006, hwang_prb_2007_lindhard,barlas_prl_2007,principi_prb_2009}.
Here we have $\Pi_0(q,0)<0$.

\begin{figure}
\begin{overpic}[width=\linewidth]{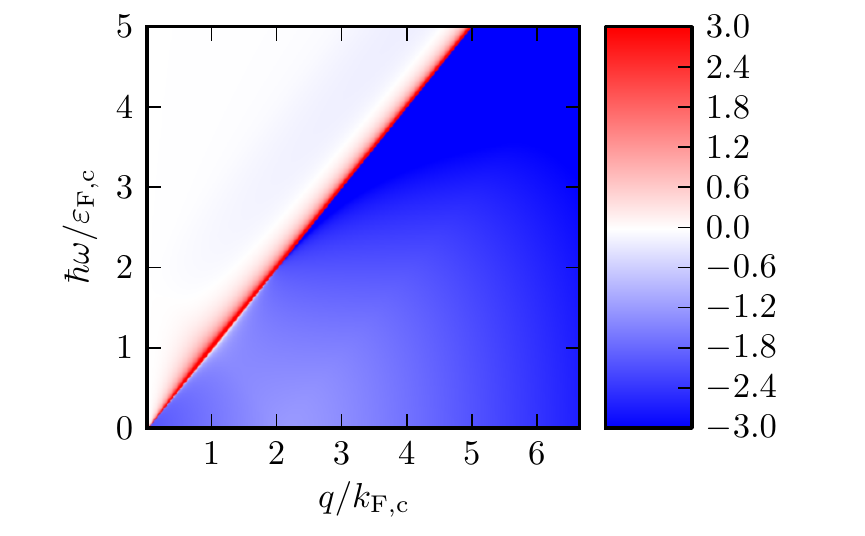}\put(2,60){(a)}\end{overpic}
\begin{overpic}[width=\linewidth]{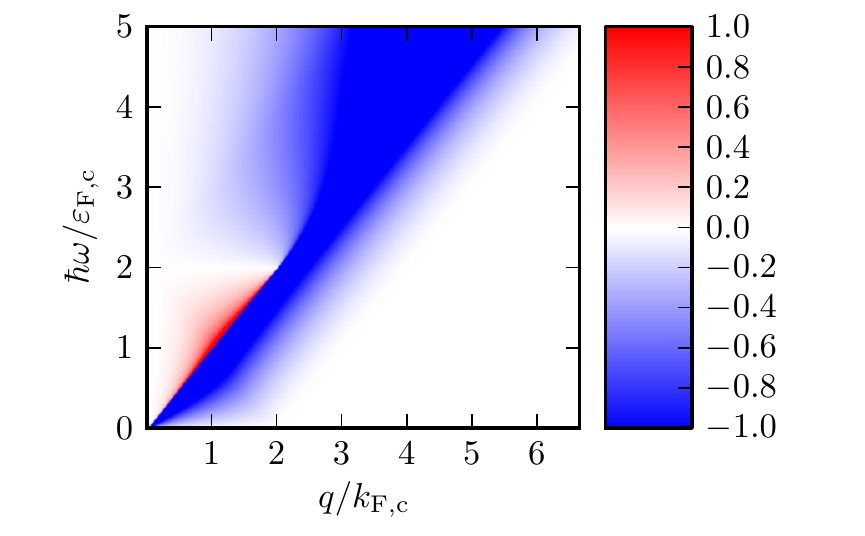}\put(2,60){(b)}\end{overpic}
\caption{(Color online) Color plots of (a) the real and (b) imaginary parts of the thermal polarization function $\chi^{(0)}(q,\omega;T)$ as calculated from Eq.~(\ref{eq:lindhardTwoMu}). 
$\Re e~[\chi^{(0)}(q,\omega;T)]$ and $\Im m~[\chi^{(0)}(q,\omega;T)]$ are plotted in units of the 2d MDF density of states $\nu(\varepsilon)= N_{\rm s} N_{\rm v} \varepsilon / (2\pi \hbar^2 v^2_{\rm F})$ evaluated at the conduction-band Fermi energy $\varepsilon = \varepsilon_{\rm F,c}$. 
Here $N_{\rm s} = N_{\rm v} = 2$ are spin and valley degeneracy factors.
These plots refer to the following parameters: $k_{\rm B} T = 0.01~{\rm eV}$, $\mu_{\rm c} = - \mu_{\rm v} = 0.6~{\rm eV}$. \label{fig:chi2D}}
\end{figure}

\begin{figure}
\begin{overpic}[width=\linewidth]{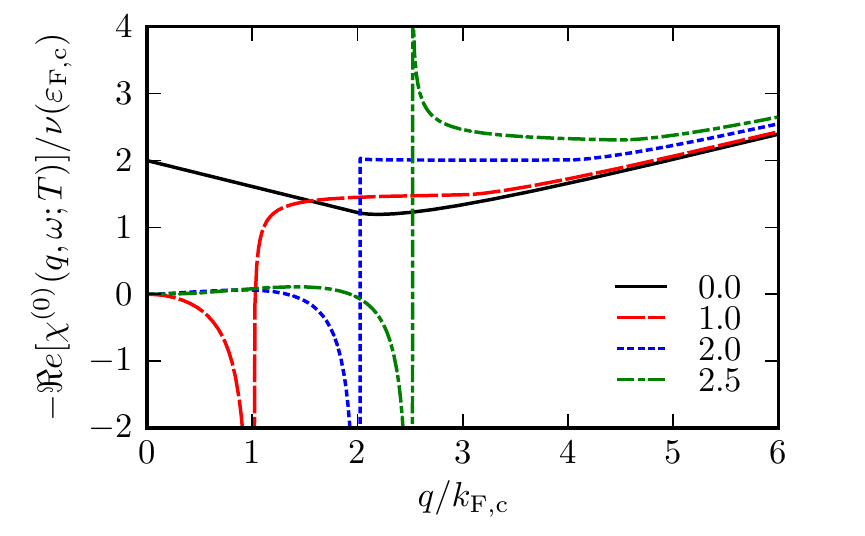}\put(2,60){(a)}\end{overpic}
\begin{overpic}[width=\linewidth]{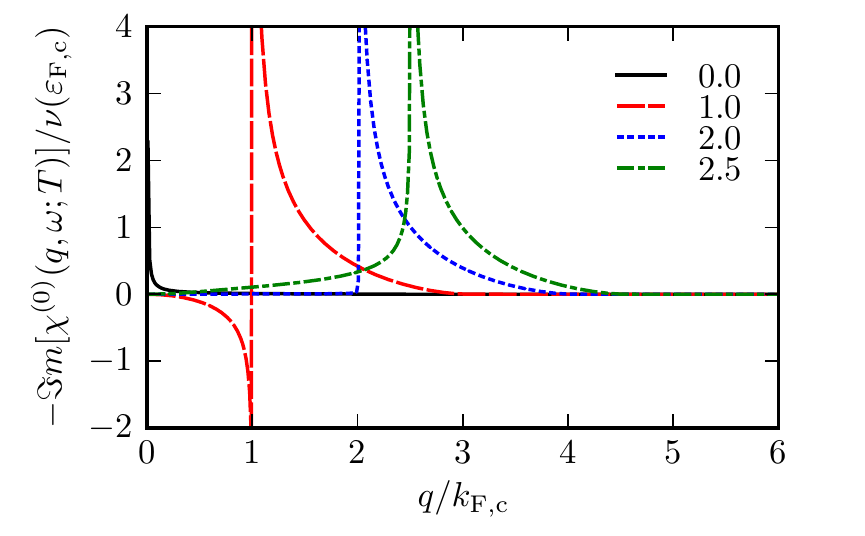}\put(2,60){(b)}\end{overpic}
\caption{(Color online) Dependence of the real, panel (a), and imaginary, panel (b), parts of the thermal polarization function $\chi^{(0)}(q,\omega;T)$ on the wave vector $q$ (in units of $k_{\rm F,c}$).
The parameters are the same as in Fig.~\ref{fig:chi2D}.
The legends show the values of $\hbar \omega / \varepsilon_{\rm F,c}$.\label{fig:chiCuts}}
\end{figure}

Eq.~(\ref{eq:lindhardTwoMu}) is the main result of this Section and reveals that the thermal polarization function $\chi^{(0)}(q,\omega; T)$ naturally decomposes into the sum of two contributions stemming from each band.
However, the extra correction in the second line of Eq.~(\ref{eq:lindhardTwoMu}) needs to be taken into account at finite $T$.
The following three identities are necessary to derive Eq.~(\ref{eq:lindhardTwoMu}): i) $F(\varepsilon_{{\bm k}, s}; \mu_{\rm c}, \mu_{\rm v}, T)=F(\varepsilon_{{\bm k}, s}; \mu_{s}, T)$, ii) $F(\varepsilon; \mu_{\rm c}, \mu_{\rm v}, 0) = \Theta(\mu_{\rm c}) F(\varepsilon;\mu_{\rm c}, 0) + \Theta(-\mu_{\rm v}) F(\varepsilon; \mu_{\rm v}, 0) + [\Theta(\mu_{\rm v}) - \Theta(\mu_{\rm c})] F(\varepsilon;0,0)$, and iii)
\begin{equation}
\frac{1}{e^x +1} = \int_{-\infty}^{+\infty}dy \frac{\Theta(y-x)}{4 \cosh^2(y/2)}~.
\end{equation}
Illustrative plots of the real and imaginary parts of the polarization function~(\ref{eq:lindhardTwoMu}) are reported in Figs.~\ref{fig:chi2D},\ref{fig:chiCuts}.
In these plots we rescaled $\hbar \omega$ with the conduction-band Fermi energy $\varepsilon_{\rm F,c} \ge 0$,
\begin{eqnarray}\label{eq:fermiEnergyConduction}
\varepsilon_{\rm F,c}  & = & k_{\rm B} T \, \sqrt{2} \sqrt{|{\rm Li}_{2}(-e^{\mu_{\rm c} / (k_{\rm B} T)})|} \nonumber \\
& \stackrel{T\to 0}{\to} & \mu_{\rm c} \, \Theta(\mu_{\rm c})~,
\end{eqnarray}
and the wave vector $q$ with the conduction-band Fermi wave number $k_{\rm F,c} = \varepsilon_{\rm F,c} / \hbar$.
In Eq.~(\ref{eq:fermiEnergyConduction}), ${\rm Li}_{2}(x)$ is the Spence's function~\cite{AbramowitzStegun}.
We note that the reactive part of the polarization function manifests a singularity along the ``light cone'' $\omega = v_{\rm F} q$, entirely inherited from $\Pi_0(q,\omega)$.
The dissipative part displays a striking difference with respect to $\Im m[\Pi_0(q,\omega)]$: the usual triangular region above the light cone where both intra- and inter-band particle-hole pairs are suppressed~\cite{kotov_rmp_2012,wunsch_njp_2006, hwang_prb_2007_lindhard,polini_prb_2008} is not present in Fig.~\ref{fig:chi2D}(b).
Moreover, above the light cone and for $\hbar \omega < 2 \mu_{\rm c}$, a region where the imaginary part of $\chi^{(0)}(q,\omega;T)$ is {\it positive} appears.

Here we approximate the time-dependent polarization function $\chi^{(0)}(q,\omega)$ in Eq.~(\ref{eq:polarization}) as:
\begin{equation}\label{eq:finalapproximation}
\chi^{(0)}(q,\omega) \approx \left.\chi^{(0)}(q,\omega;T)\right|_{\mu_s \to \mu_s(t); T \to T(t)}~.
\end{equation}

\subsection{Screening models for the semiclassical Boltzmann equation}
\label{ssec:screeningModels}

In this Article we focus on the following three screening models:

i) {\it Dynamical screening}.
In this case the polarization function $\chi^{(0)}(q,\omega;T)$ in Eq.~(\ref{eq:finalapproximation}) is evaluated ``on shell", i.e.~at $\hbar \omega = \varepsilon_{1} - \varepsilon_{3}$, which is the energy transferred in the two-body scattering process.
As mentioned above, in this case inter-band and intra-band scattering rates are finite while the Auger contribution (\ref{eq:AugerResult}) to the Coulomb kernel vanishes.

ii) {\it Static screening}.
In  this case the polarization function $\chi^{(0)}(q,\omega;T)$ in Eq.~(\ref{eq:finalapproximation}) is evaluated at $\omega = 0$.
This approximation is justified when the energy $\hbar \omega = \varepsilon_{1} - \varepsilon_{3}$ transferred in the two-body scattering process is significantly smaller than the energy $\Omega_{\rm pl}(|{\bm k}_1 - {\bm k}_3|)$, necessary to excite a plasmon~\cite{wunsch_njp_2006,hwang_prb_2007,polini_prb_2008}.
In the static screening approximation, inter-band and intra-band scattering rates diverge and can be regularized by employing an infrared and an ultraviolet cutoff~\cite{fritz_prb_2008} (further details are reported in the Appendix).
We find, however, that the resulting non-equilibrium dynamics does not depend on the values of these two cutoffs.
The reason is the following.
Intra- and inter-band scattering processes are responsible only for redistributing energy, rapidly driving the two bands towards thermal equilibrium.
Therefore, for $\varepsilon_1$ and $\varepsilon_3$ varying in regions of type I and II (see Fig.~\ref{fig:IntegrationDomain}) the quantity in curly brackets in Eq.~(\ref{eq:finalboltzmann}) vanishes.
In the static approximation the Auger contribution (\ref{eq:AugerResult}) to the Coulomb kernel is finite.

iii) {\it Regularized screening}.
Finally, we introduce a ``regularized'' screening model~\cite{brida_arxiv_2012} in which the polarization function $\chi^{(0)}(q,\omega;T)$ in Eq.~(\ref{eq:finalapproximation}) is evaluated on shell, but its singularity around the light cone $\omega = v_{\rm F} q$ is {\it smeared} by means of a cutoff $\Lambda_{\rm E}$ (see App.~\ref{ssec:numerical} for more details).
In the limit $\Lambda_{\rm E} \to 0$ this model reduces to dynamical screening, as described at point i) above.
Various physical mechanisms can smear the singularity of the polarization function on the light cone, including many-body effects beyond RPA (as suggested in Ref.~\onlinecite{brida_arxiv_2012}) or single-particle effects beyond the 2d MDF model, e.g.~trigonal warping (as suggested in Ref.~\onlinecite{basko_prb_2013}).
In our regularized screening model the Auger contribution (\ref{eq:AugerResult}) to the Coulomb kernel is finite and, in particular, its magnitude is intermediate between that evaluated within the dynamical and static screening models.

\section{Numerical results}
\label{sec:Results}

In this Section we summarize our main numerical results, obtained from the solution of the isotropic SBE within the three screening models listed in Sec.~\ref{ssec:screeningModels}.

We compared the outcome of a closely related approach with our experimental measurements in Ref.~\onlinecite{brida_arxiv_2012}.
There we applied screening to both direct and exchange terms in Eq.~(\ref{eq:HF}).
The data that we present in this Section are in agreement with our previous results in Ref.~\onlinecite{brida_arxiv_2012}, therefore fully support the interpretation of the experimental data given there.
More precisely, very good quantitative agreement is found for the static and dynamical screening models, which do not depend on free parameters.
There are some differences, however, between the dynamical behavior illustrated in Ref.~\onlinecite{brida_arxiv_2012} and that discussed in this Section, when the regularized screening model is used.
Although the mathematical definition of the cutoff $\Lambda_{\rm E}$ here is analogous to the definition of $\Lambda$ in Ref.~\onlinecite{brida_arxiv_2012}, the equations of motion induced by the regularized screening models differ.
As repeatedly stressed above, in the present paper we never screen dynamically the exchange contribution to Eq.~(\ref{eq:HF}).
Moreover, here we mostly focus on undoped samples, although some results for $n$-doping are presented in Fig.~\ref{fig:maximumTransmission}.
On the other hand, Ref.~\onlinecite{brida_arxiv_2012} reported experiments and calculations for a $p$-doped sample.
Even though Ref.~\onlinecite{brida_arxiv_2012} already allowed us to conclude that RPA dynamical screening is not capable of explaining the experimental results, in this Article for sake of completeness we will present a comparative study of all screening models listed in Sec.~\ref{ssec:screeningModels}.

\subsection{Choice of the initial hot-electron distribution}
\label{initialstate}

As initial condition for the solution of the SBE we use a distribution function which is the sum of a FD distribution, with chemical potential $\mu$ and temperature $T_0$, and two Gaussian-shaped peaks (one below and one above the Dirac point):
\begin{eqnarray}\label{eq:initialstate}
\left.f_\ell(\varepsilon)\right|_{t =0}  &=& F(\varepsilon;\mu,T_{0}) + g_{\rm max} \exp{\left[-\left(\frac{\varepsilon - \hbar \omega_{\rm P}/2}{\hbar \Delta \omega_{\rm P}/2}\right)^2\right]}\nonumber\\
&-& g_{\rm max} \exp{\left[-\left(\frac{\varepsilon + \hbar \omega_{\rm P}/2}{\hbar \Delta \omega_{\rm P}/2}\right)^2\right]}~.
\end{eqnarray}
The initial distribution function is identical in both $\ell = {\rm K}, {\rm K}'$ valleys.

The choice (\ref{eq:initialstate}) is motivated by pump-probe spectroscopy experiments on graphene, where electrons are promoted from valence to conduction band by using a laser light pulse.
Before the pump pulse is applied, the electronic subsystem is at equilibrium with the lattice at a given $T$, say $T_0 = 300~{\rm K}$ (room $T$).
The momentum transferred by the laser light to the electrons is negligible, hence the transitions are ``vertical'' in momentum space, from energy $\varepsilon = -\hbar\omega_{\rm P}/2$ in valence band to energy $\varepsilon = \hbar\omega_{\rm P}/2$ in conduction band.
The width $\hbar \Delta \omega_{\rm P}$ of the light pulse determines the width of the resulting HED.
Consistent with our recent experiments~\cite{brida_arxiv_2012}, and for the sake of definiteness, we take $\hbar \omega_{\rm P} = 2.25~{\rm eV}$, $\hbar \Delta \omega_{\rm P}/2 = 0.09~{\rm eV}$, and $g_{\rm max} = 0.5$.
We point out that these parameters correspond to a strongly non-equilibrium distribution, obtained by shining a light pulse with fluence $\gtrsim ~\mu{\rm J}/{\rm cm}^{-2}$.

Finally, we note that, although in general the light-matter coupling is anisotropic, the HED has been shown to relax to an isotropic profile in $\simeq 5~{\rm fs}$~\cite{malic_prb_2011}.
As we mentioned above, the description of these very early stages of the non-equilibrium dynamics, which comprise the buildup of the HED, is beyond the scope of the present Article.
Here, we study the time evolution of the isotropic initial state (\ref{eq:initialstate}), as dictated by the SBE.

\subsection{Values of the electron-optical phonon coupling constants}

The energies of the phonon modes are~\cite{BPF2009} $\hbar \omega^{(\Gamma)} \simeq 0.150~{\rm eV}$ and $\hbar \omega^{({\rm K})} \simeq 0.196~{\rm eV}$.
The EPC of the $E_{2g}$ phonon at the Brillouin-zone center ($\Gamma$ point), associated to the $G$ peak of the Raman spectrum, is taken from Ref.~\onlinecite{PLMFR2004}: $\langle g_{\Gamma}^{2} \rangle \simeq 0.0405~{\rm eV}^{2}$.
This value, which we use in our numerical calculations, is in good agreement with experimental results~\cite{BPF2009}.
On the other hand, the value of the EPC relative to the transverse mode at the ${\rm K}$ point has been debated~\cite{BPF2009}.
The value calculated by density-functional theory~\cite{PLMFR2004} is $\langle g_{{\rm K},2}^{2} \rangle \simeq 0.0994~{\rm eV}^{2}$, but e-e interactions renormalize this value by a factor $2-5$~\cite{ferrari_prl_2006,gruneis_prb_2009,berciaud_nanolett_2009,BPF2009}.

In our numerical calculations we take $\langle g_{{\rm K},2}^{2} \rangle \simeq 0.2~{\rm eV}^{2}$. The EPC of the longitudinal phonon mode at ${\rm K}$ point is taken to be $\langle g_{{\rm K},1}^{2} \rangle \simeq 0.00156~{\rm eV}^{2}$, as in Ref.~\onlinecite{butscher_apl_2007}.
Following Ref.~\onlinecite{sun_prb_2012}, we take $\gamma_{\rm ph} / \hbar \simeq 0.26~{\rm ps}^{-1}$ for the phenomenological phonon decay rate in Eq.~(\ref{eq:phonon}).

\subsection{Role of Auger scattering}
\label{ssect:ResultsAuger}

\begin{figure}
\begin{overpic}[width=\linewidth]{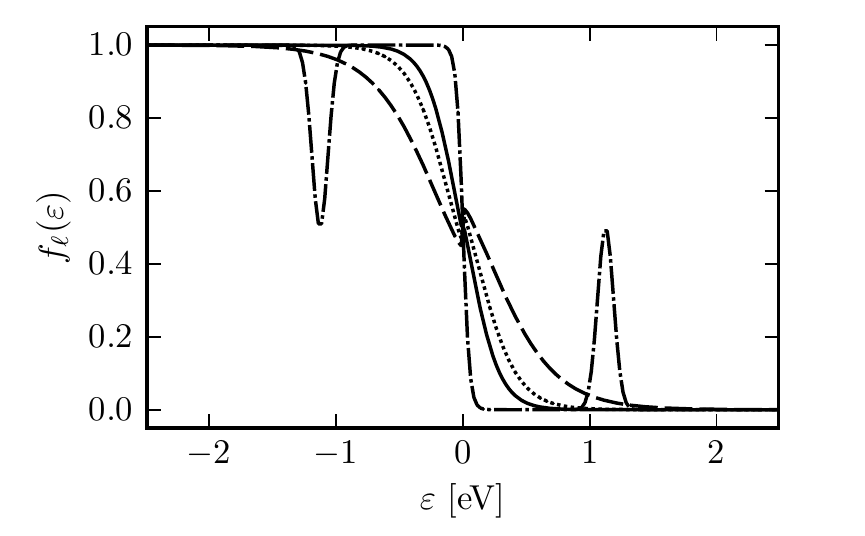}\put(2,60){(a)}\end{overpic}
\begin{overpic}[width=\linewidth]{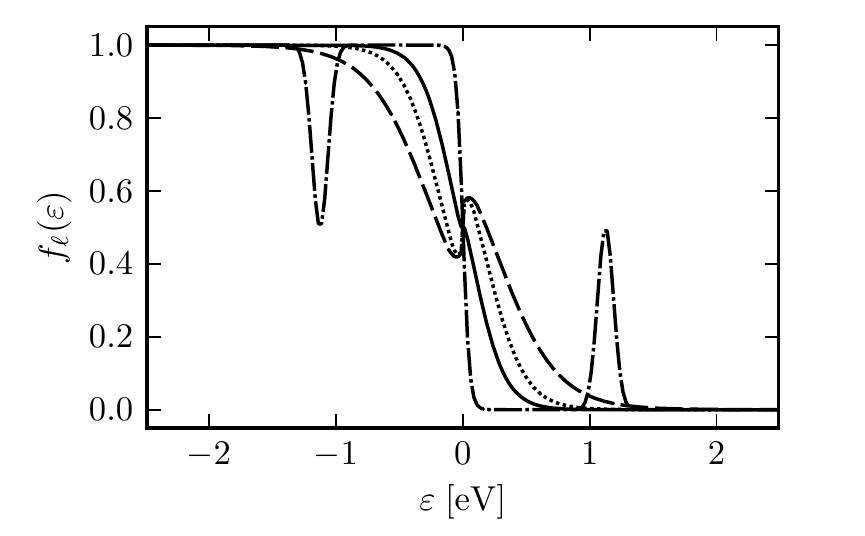}\put(2,60){(b)}\end{overpic}
\begin{overpic}[width=\linewidth]{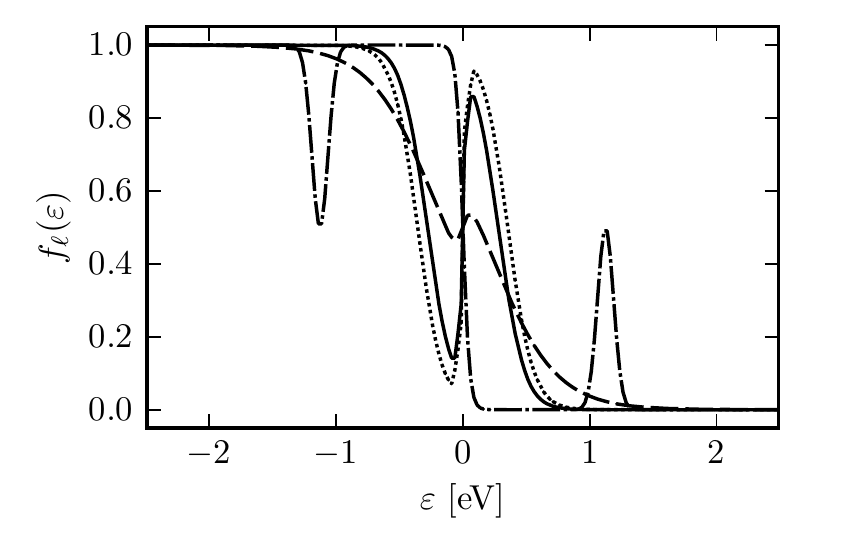}\put(2,60){(c)}\end{overpic}
\caption{Time evolution of the electron distribution function $f_\ell(\varepsilon)$ as dictated by the isotropic SBE with: static screening, panel (a), regularized dynamical screening, panel (b), and dynamical screening, panel (c). In all panels different line styles refer to three different times: $t = 0$ (dash-dotted line), $t = 100.0~{\rm fs}$ (dashed line), $t = 500.0~{\rm fs}$ (dotted line), and $t = 1.0~{\rm ps}$ (solid line). The creation of a large inverted carrier population around the Dirac point $\varepsilon = 0$ is seen in panel (c). \label{fig:evolutionDistro}}
\end{figure}
\begin{figure}
\begin{overpic}[width=\linewidth]{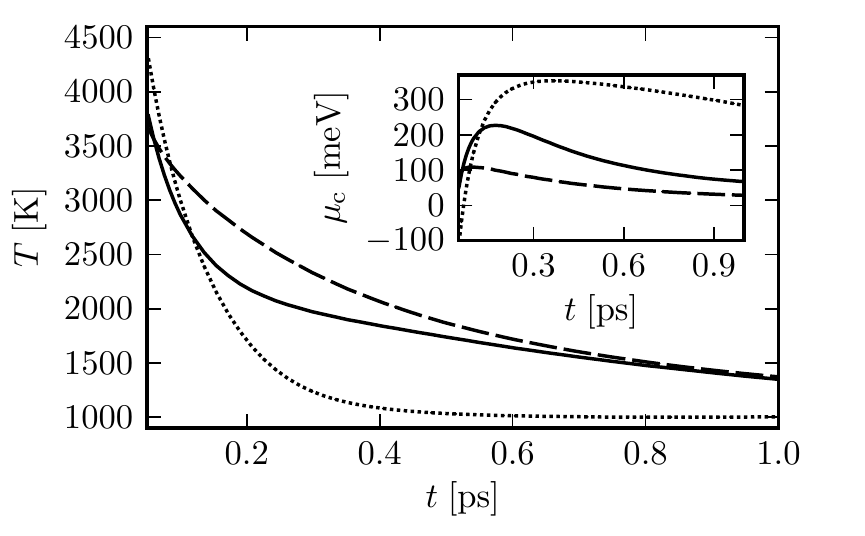}\put(2,60){(a)}\end{overpic}
\begin{overpic}[width=\linewidth]{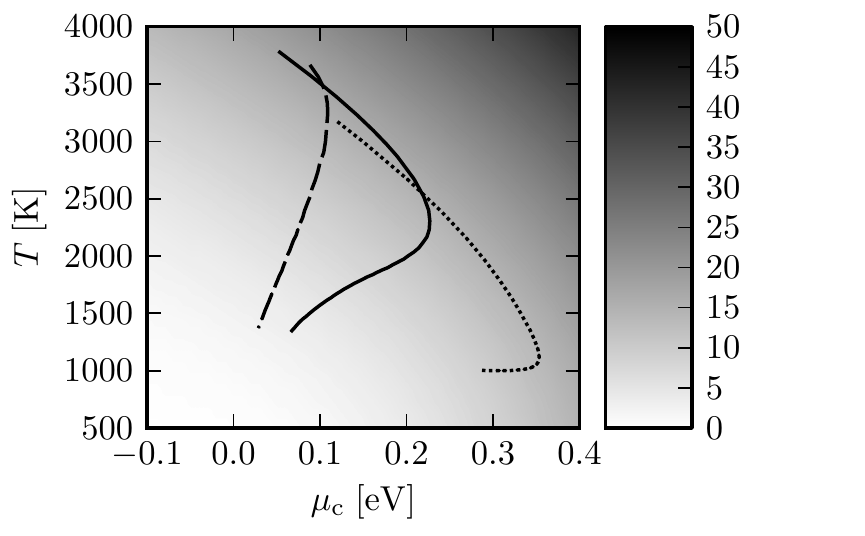}\put(2,60){(b)}\end{overpic}
\caption{(a) Time evolution of $T$ (in ${\rm K}$) and chemical potential $\mu_{\rm c}$ (in ${\rm meV}$) in conduction band (inset). Different line styles refer to the three screening models: static (dashed line), regularized (solid line), and dynamical (dotted line).
Dynamical screening reduces $T$ at the expense of a much larger $\mu_{\rm c}$.
(b) Color plot of the electron density in conduction band (in units of $10^{12}~{\rm cm}^{-2}$). The lines show the relation between $T$ and $\mu_{\rm c}$ during the time evolution [line styles as in panel (a)].
Note that, in the presence of dynamical screening, a much longer stage of the time evolution exists in which no loss of electrons from conduction band takes place.\label{fig:thermalization}}
\end{figure}

We start by discussing the role of different screening models, and choose a strength of e-e interactions $\alpha_{\rm ee} = 0.9$, appropriate~\cite{kotov_rmp_2012} for graphene on a SiO$_{2}$ substrate, see Eq.~(\ref{eq:FSC}).
The corresponding SBE solution for undoped graphene is shown in panels (a)--(c) of Fig.~\ref{fig:evolutionDistro} for static, regularized, and dynamical screening, respectively.
In all cases we see that the peak (dip) of the HED above (below) the Dirac point shifts rapidly towards the Dirac point.
There is however a striking difference between panels (a),(b) and panel (c): in dynamical screening, panel (c), a much larger electron (hole) population persists in conduction (valence) band even at times as long as $t = 1.0~{\rm ps}$. The reason is that dynamical screening suppresses AR events, thus delays equilibration of the electron populations across the two bands.
Indeed, in the initial stage of the time evolution, Auger processes are the most important processes for the equilibration of the electron populations.
On a longer time scale, relaxation by phonon emission allows the system to reach inter-band equilibrium.
However, in this case, the existence of a substantial inverted carrier population around the Dirac point at times as long as $1~{\rm ps}$ is due to the suppression of AR processes.
We point out that thermal equilibrium between the two bands, on the contrary, mainly occurs {\it via} inter-band scattering, and is reached after a much shorter time $\simeq 20~{\rm fs}$.

A more quantitative analysis of the inter-band equilibration dynamics is shown Fig.~\ref{fig:thermalization}.
Here we also report numerical results based on the regularized screening model.
A FD distribution with time-dependent temperature $T(t)$ and chemical potential $\mu_{\rm c}(t)$ [$\mu_{\rm v}(t)$] in conduction (valence) band can be fitted to the numerical results from the solution of the isotropic SBE for $t > 20~{\rm fs}$. $T(t)$ remains well above room $T$ for $t \lesssim 1~{\rm ps}$.
In the absence of AR processes, a much faster cooling of the initial HED occurs, at the price of a larger chemical potential $\mu_{\rm c}(t)$ in conduction band.
The energy stored in the electronic degrees of freedom is then transferred to the phonon modes and dissipated into the lattice by means of phonon-phonon interactions, responsible for the phenomenological decay term proportional to $\gamma_{\rm ph}$ in Eq.~(\ref{eq:phonon}).
Eventually, equilibration with the lattice at room $T(t)$ is achieved (data not shown). Figs.~\ref{fig:evolutionDistro}, \ref{fig:thermalization} indicate that different screening models strongly affect the HED time evolution.
\begin{figure}
\includegraphics[width=\linewidth]{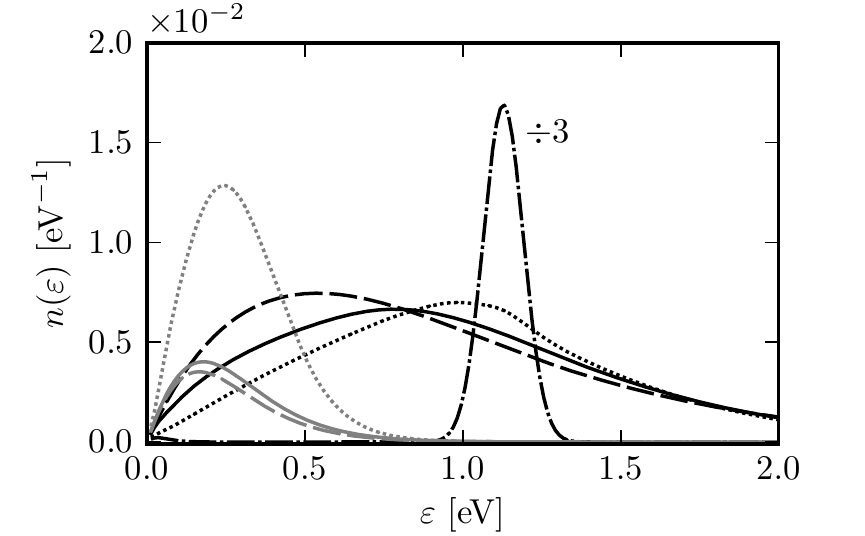}
\caption{Electron density $n(\varepsilon)$ per unit cell area and energy obtained from the numerical solution of the isotropic SBE as a function of energy $\varepsilon$ (in units of ${\rm eV}$).
Black curves refer to $t = 12.0~{\rm fs}$.
Gray curves to $t = 1.0~{\rm ps}$.
Different line styles refer to the three screening models: static (dashed lines), regularized (solid lines), and dynamical (dotted lines).
The initial state at $t = 0$ (dash-dotted line) was divided by a factor $3$ to fit into the frame of the figure.\label{fig:evolutionDensity} }
\end{figure}
\begin{figure}
\includegraphics[width=\linewidth]{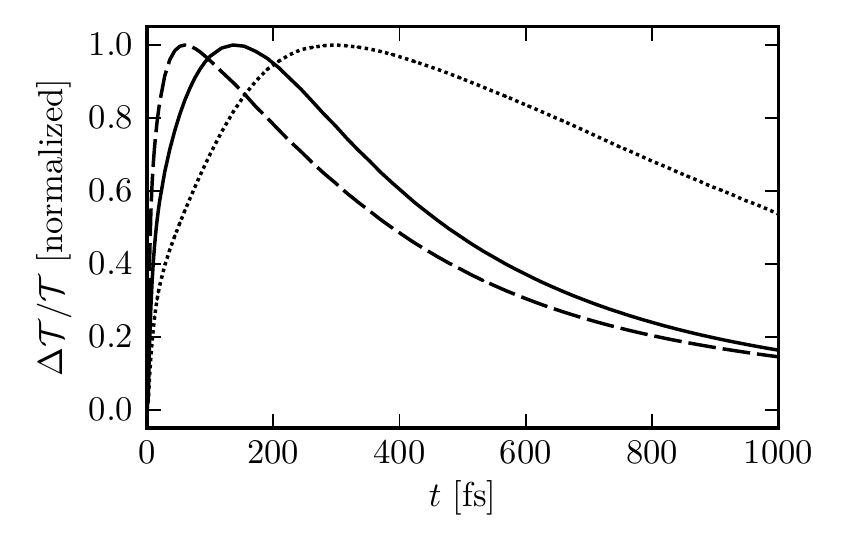}
\caption{\label{fig:transmission} Time evolution of the differential transmission $\Delta {\cal T} / {\cal T}$ as calculated from the numerical solution of the SBE.
Different line styles refer to the three screening models: static (dashed line), regularized (solid line), and dynamical (dotted line).
The data in this figure refer to a probe energy $\hbar \omega_{\rm p} = 0.8~{\rm eV}$.
Each curve is normalized to have maximum $\Delta {\cal T} / {\cal T} |_{\rm max} = 1.0$.
Note that dynamical screening gives a much slower time evolution since it completely suppresses Auger scattering.}
\end{figure}
In Fig.~\ref{fig:evolutionDensity} we illustrate the dependence of the electron density per unit cell area and energy,
\begin{equation}\label{densityversusenergy}
n(\varepsilon) = {\cal A}_{0} \nu(\varepsilon) f_{\ell}(\varepsilon) \simeq \varepsilon \, f_{\ell}(\varepsilon) \times 0.09~{\rm eV}^{-2}~,
\end{equation}
on the energy $\varepsilon$. Note that $n(\varepsilon)$ has dimensions ${\rm eV}^{-1}$. In Eq.~(\ref{densityversusenergy})
\begin{equation}
\nu(\varepsilon) = \frac{N_{\rm s} N_{\rm v} \varepsilon}{2\pi \hbar^2 v^2_{\rm F}} \simeq \varepsilon \times 1.77~{\rm nm}^{-2} {\rm eV}^{-2}
\end{equation}
is the 2d MDF density of states as a function of energy.
The quantity $N_{\rm v} = 2$ represents the valley degeneracy.
The energy at which $n(\varepsilon)$ peaks strongly depends on the screening model.
The optical properties of the MDF system are very sensitive to the time evolution of $n(\varepsilon)$ since light absorption is strongly inhibited (Pauli blocking) when the corresponding electronic transitions are towards states with a larger occupation.
By shining a probe laser pulse with frequency $\omega_{\rm p}$ through the sample one can measure the time evolution of the electron distribution, a procedure enabled by Pauli blocking.
A viable experimental route to directly measure the impact of screening is thus available, provided that short enough probe pulses of appropriate frequency are used~\cite{brida_arxiv_2012}.

The propagation of the probe pulse through the sample can be quantified by calculating the differential transmission~\cite{breusing_prb_2011} (DT)
\begin{eqnarray}
\frac{\Delta {\cal T}}{{\cal T}}(\omega_{\rm p},t) & = & \pi \alpha [ f_{\ell}(\hbar \omega_{\rm p} / 2) - F(\hbar \omega_{\rm p} / 2; \mu, T_0) \nonumber \\
& & - f_{\ell}(-\hbar \omega_{\rm p} / 2) + F(-\hbar \omega_{\rm p} / 2; \mu, T_0)]~, \nonumber \\
\end{eqnarray}
where $\alpha = e^{2} / (\hbar c) \simeq 1/137$ is the fine-structure constant and $\mu$ and $T_0$ are the chemical potential and temperature of the electron system before the pump pulse is applied.
The time evolution of the normalized DT is shown in Fig.~\ref{fig:transmission} for the three screening models, at a fixed value of the probe energy $\hbar\omega_{\rm p}$.
The much slower dynamics in the absence of Auger processes (dynamical screening, dotted line) is clearly seen.
The time $t_{\rm max}$ at which the DT peaks is a convenient measure of the speed of the electron dynamics.
Below we discuss the dependence of $t_{\rm max}$ on various relevant parameters, see Fig.~\ref{fig:maximumTransmission}.
\begin{figure}
\includegraphics[width=\linewidth]{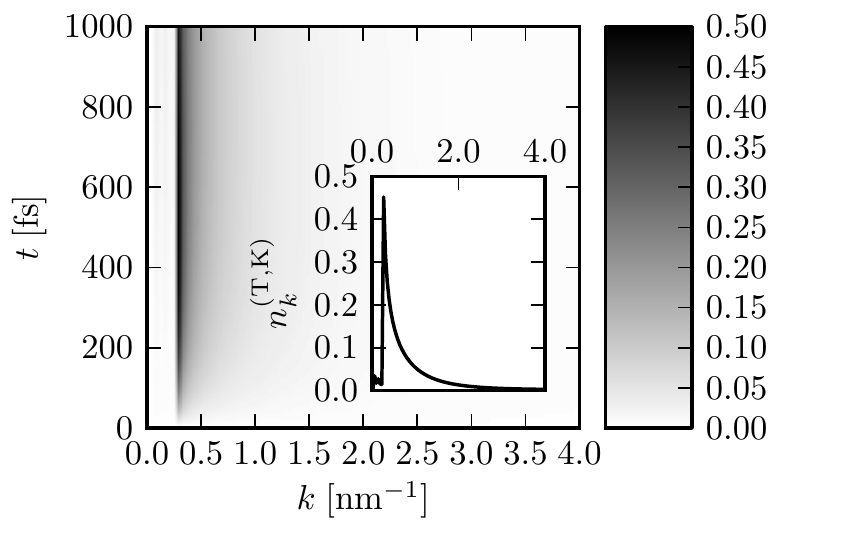}
\caption{Color plot of the phonon distribution $n^{(\nu)}_{k}(t)$ as a function of wave vector $k$ (in ${\rm nm}^{-1}$) and time $t$ (in units of ${\rm fs}$), showing the distribution function $n^{(\nu)}_{k}(t)$ for the transverse phonon mode at ${\rm K}$ ($\nu = {\rm T},{\rm K}$).
The inset shows the same quantity at $t = 1.0~{\rm ps}$.\label{fig:evolutionPhonon}}
\end{figure}
Although the focus of the present Article is on the time evolution of the electron distribution function, for the sake of completeness in Fig.~\ref{fig:evolutionPhonon} we illustrate the time evolution of the distribution function of the transverse optical phonon mode at ${\rm K}$, i.e.~the mode most strongly coupled to the electronic subsystem~\cite{PLMFR2004,lazzeri_prl_2005,sun_acsnano_2010}.
In a sub-$100~{\rm fs}$ time a large population accumulates in the mode (with respect to the equilibrium population) and remains steady up to the maximum time $t = 1.0~{\rm ps}$ considered here.
This hot phonon distribution~\cite{lazzeri_prl_2005} cools on a longer time scale as outlined in Sec.~\ref{ssec:timescales} by anharmonic phonon-phonon interactions, which dissipate the thermal energy into the lattice.

Thus, electronic screening is responsible for qualitative modifications of the dynamics in the sub-$100~{\rm fs}$ time scale, which were not unveiled previously to the best of our knowledge.
This early stage of the dynamics is temporally decoupled from other relaxation channels (phonon and radiative emission), thus the effects of electronic screening do not modify the hot-electron relaxation picture as outlined in Sec.~\ref{ssec:timescales}.
At later times, our numerical results broadly agree with previous theoretical works~\cite{butscher_apl_2007,winzer_nanolett_2010,malic_prb_2011,kim_prb_2011,girdhar_apl_2011,sun_prb_2012,winzer_prb_2012, winzer_jpcm_2013,sun_arxiv_2013}.

\subsection{Role of EPC, doping, e-e interaction strength, and exchange}
\label{ssect:ResultsOther}

\begin{figure}
\begin{overpic}[width=\linewidth]{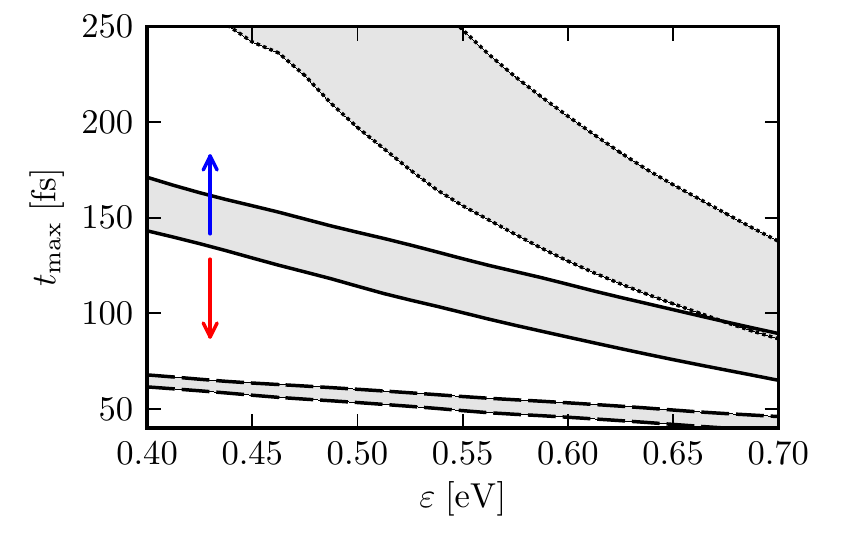}\put(2,60){(a)}\end{overpic}
\begin{overpic}[width=\linewidth]{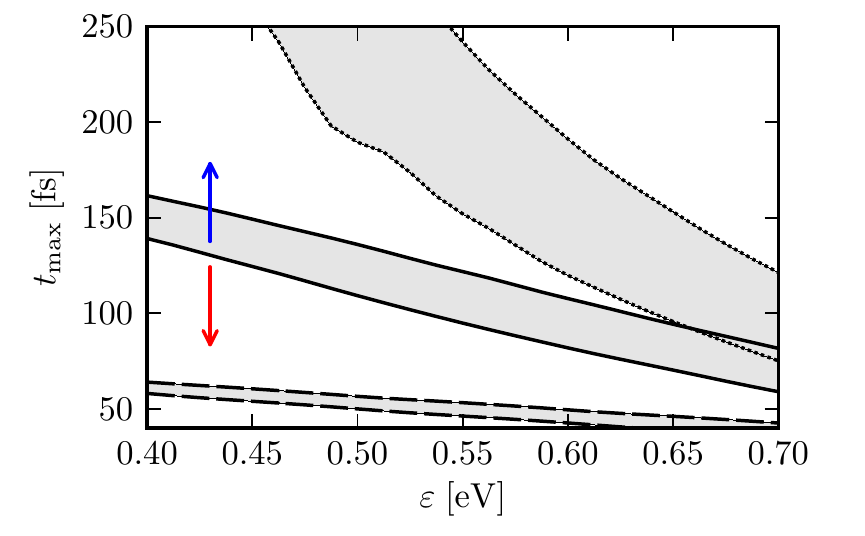}\put(2,60){(b)}\end{overpic}
\caption{\label{fig:maximumTransmission}
(Color online) $t_{\rm max}$ (in ${\rm fs}$) as a function of electron energy $\varepsilon$ (in ${\rm eV}$).
Panel (a) consider an undoped sample, while panel (b) a $n$-doped one with a chemical potential $\mu = 200~{\rm meV}$.
Both panels refer to a lattice temperature $T_0 = 300~{\rm K}$.
Results for static (dashed lines), regularized (solid lines), and dynamical (dotted lines) screening are shown.
The lower (upper) border of the gray-shaded regions is evaluated by choosing $\langle g_{{\rm K},2}^{2} \rangle = 0.0994~{\rm eV}$ ($0.2~{\rm eV}$).
Note that the dynamics resulting from the three different screening models is substantially different in a wide range of EPCs.
These results agree with Fig.~5(c) in Ref.~\onlinecite{brida_arxiv_2012}, where numerical and experimental results for a $p$-doped system were presented.
As explained in Sec.~\ref{ssec:screeningModels}, the results obtained with regularized screening are intermediate between static and dynamical, and depend on the magnitude of the cutoff $\Lambda_{\rm E}$, as indicated by the arrows (red arrow: $\Lambda_{\rm E}$ increases; blue arrow: $\Lambda_{\rm E}$ decreases).
The regularized screening model cannot be quantitatively compared with Fig.~5(c) in Ref.~\onlinecite{brida_arxiv_2012} because the $\Lambda_{\rm E}$ values cannot be simply mapped to those of $\Lambda$ in Ref.~\onlinecite{brida_arxiv_2012}, as explained at the beginning of Sec.~\ref{sec:Results}.}
\end{figure}

\begin{figure}
\begin{overpic}[width=\linewidth]{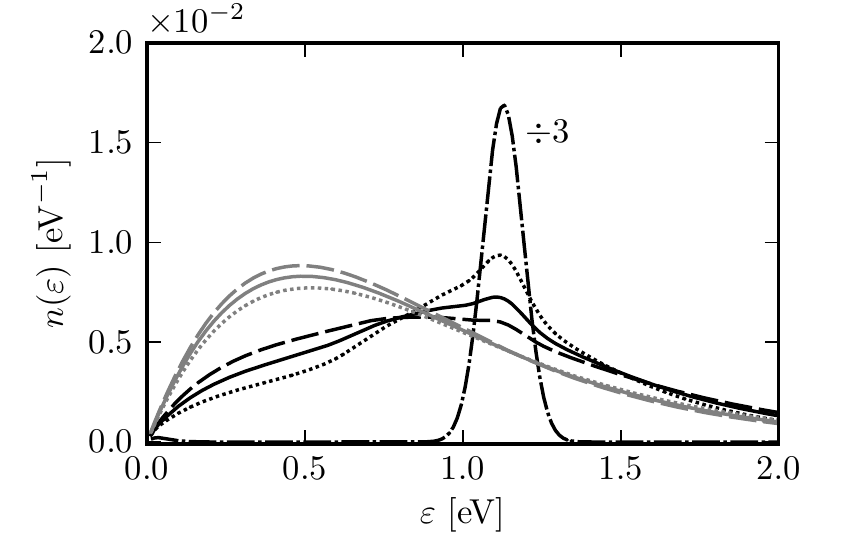}\put(2,60){(a)}\end{overpic}
\begin{overpic}[width=\linewidth]{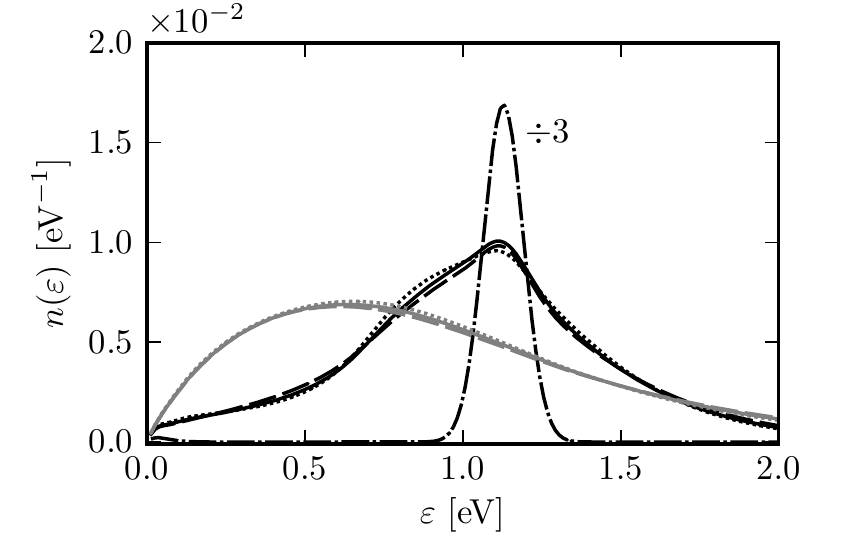}\put(2,60){(b)}\end{overpic}
\caption{Electron density per unit lattice cell and energy as obtained from the numerical solution of the SBE with two screening models: static, panel (a), and regularized dynamical, panel (b).
Data in this figure refer to $t = 4.0~{\rm fs}$ (black curves) and $t = 18.0~{\rm fs}$ (gray curves).
Different line styles refer to three values of the graphene's fine-structure constant: $\alpha_{\rm ee} = 0.5$ (dotted lines), $\alpha_{\rm ee} = 0.9$ (solid lines), and $\alpha_{\rm ee} = 2.2$ (dashed lines).
The initial state at $t = 0$ (dash-dotted line) is divided by a factor $3$ to fit into the frames of the panels.\label{fig:effectsEE}}
\end{figure}

We now show that the results presented in the previous Section are robust with respect to changes in parameter space.
We find that, to a large extent, the speed of the relaxation dynamics in the sub-$100~{\rm fs}$ time range is controlled by the particular screening model one chooses.

Fig.~\ref{fig:maximumTransmission}(a) plots $t_{\rm max}$ as a function of $\varepsilon = \hbar \omega_{\rm p}/2$ for two choices of the largest EPC, $\langle g_{{\rm K},2}^{2} \rangle$, gray-shading the region in-between.
The largest EPC (then used in all other figures) yields the smallest $t_{\rm max}$, i.e.~a faster relaxation dynamics.
This is due to the fact that a larger coupling of the electrons to the phonon bath allows a more efficient dissipation of the excess energy.
The DT peaks later for smaller energy, reflecting the shifting of the peak of the electron density $n(\varepsilon)$ towards the Dirac point (see Fig.~\ref{fig:evolutionDensity}).
The three screening models give quantitatively different results, with static screening remaining in the sub-$100~{\rm fs}$ range, and dynamical screening showing a much more pronounced dependence on electron energy.
Most importantly, there is very limited overlap between gray-shaded regions, meaning that the three screening models yield distinctly different relaxation speeds, even if the EPC is increased by a factor $2$.
These results are robust with respect to doping: Fig.~\ref{fig:maximumTransmission}(b), e.g., illustrates essentially unchanged results for an $n$-doped sample, with a finite positive chemical potential $\mu = 200~{\rm meV}$.

Fig.~\ref{fig:effectsEE} compares the time evolution of the electron density $n(\varepsilon)$ as calculated for $\alpha_{\rm ee} = 0.9$, with results obtained for $\alpha_{\rm ee} = 0.5$ (describing graphene on hexagonal boron nitride~\cite{dean_naturenanotech_2010}) and $\alpha_{\rm ee}=2.2$, the maximum value corresponding to a suspended graphene~\cite{grigorenko_naturephoton_2012}.
As expected, the broadening of the initial photo-excited electron distribution is faster for larger $\alpha_{\rm ee}$ (due to enhanced Coulomb repulsion) but, in general, $n(\varepsilon)$ at a given $t$ depends weakly on the e-e interaction strength.
Using regularized dynamical screening instead of static further reduces the effects of increasing $\alpha_{\rm ee}$. This behavior can be understood by recalling that electron thermalization takes place in a sub-$20~{\rm fs}$ time interval, as pointed out earlier~\cite{malic_prb_2011,sun_prb_2012,brida_arxiv_2012}.
This means that the electron distribution rapidly reaches the form of a quasi-equilibrium FD distribution which nullifies the intra- and inter-band contributions of the collisional integral (Eq.\ref{eq:finalboltzmann}), as remarked in Sec.~\ref{ssec:screeningModels} as well.
Variations in the the e-e coupling constant, which multiplies the collisional integral, contribute minor corrections to the dynamics in the collisional regime.
\begin{figure}
\includegraphics[width=\linewidth]{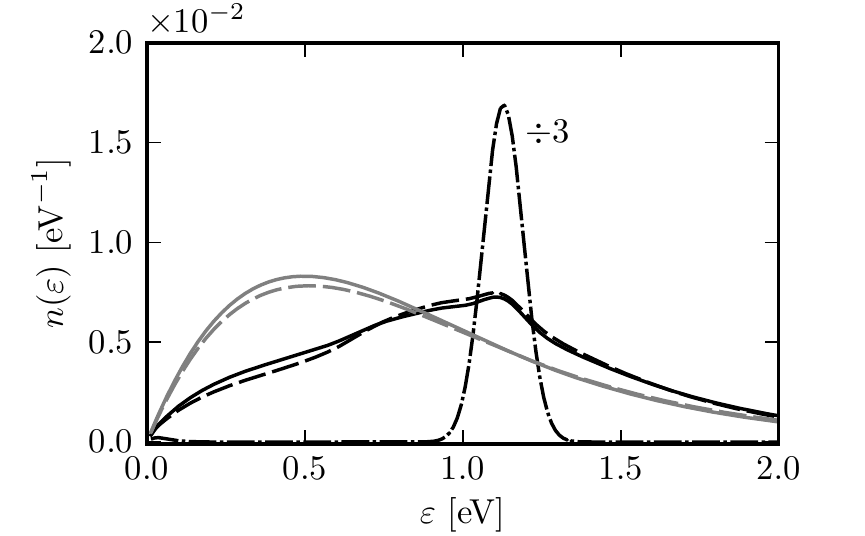}
\caption{Electron density per unit lattice cell and energy as obtained from the numerical solution of the SBE with $\alpha_{\rm ee} = 0.9$ and {\it static screening}.
Black (gray) lines refer to $t = 4.0~{\rm fs}$ ($t = 18.0~{\rm fs}$).
Solid (dashed) lines refer to results without (with) the exchange contribution to the scattering amplitude, second term in Eq.~(\ref{eq:HF}).
The initial state at $t = 0$ (dash-dotted line) is divided by a factor $3$ to fit into the frame of the figure.\label{fig:effectsExchange}}
\end{figure}

Finally, Fig.~\ref{fig:effectsExchange} shows the role of the exchange term $V_{1243}^{(\mu)}$ in Eq.~(\ref{eq:HF}), by treating screening statically to ensure a particle-number-conserving collision integral.
In this case $\varepsilon(q,0)$ is calculated at $q = |{\bm k}_{1} - {\bm k}_{3}|$ for the direct term and at $q = |{\bm k}_{1} - {\bm k}_{4}|$ for the exchange term.
As clearly seen in Fig.~\ref{fig:effectsExchange}, the exchange term is responsible for small corrections to the electron density $n(\varepsilon)$, although it is expected~\cite{QV2005} that larger corrections may arise in a Fermi liquid when perturbations are applied close to the Fermi surface.

\section{Discussion}
\label{sec:Conclusions}

We studied the non-equilibrium dynamics of a high-density photo-excited electron distribution in graphene.
We used the massless Dirac Fermion model and a semiclassical Boltzmann equation approach, which includes electron and optical phonon degrees of freedom.
Our approach neglects light-matter interactions in the very early stages of the dynamics (before the electron distribution becomes isotropic), quantum effects (coherences), and non-Markovian memory effects.
A quantitative analysis of these effects is beyond the scope of this work. 

Taking into consideration light-matter interactions allowed us to describe the buildup of an anisotropic HED.
Ref.~\onlinecite{malic_prb_2011} calculated that the electron distribution evolves to an isotropic profile on a 10-fs timescale.
We therefore decided for computational convenience to consider an isotropic distribution as initial condition for the semiclassical Boltzmann equation, without loss of generality.
We also assumed the pump pulse, which creates the initial HED, to be sufficiently short to neglect phonon-scattering-induced dephasing.
Indeed, it was experimentally shown~\cite{leitenstorfer_prl_1994} that the latter effect should be considered when applying pump pulses with duration comparable to the electron-phonon scattering time, which exceeds the $100~{\rm fs}$ time scale of our present work.
The effect of phonon-scattering-induced dephasing is to broaden the initial HED.
We checked that our results are very stable with respect to changes of this type.
The Coulomb-scattering-induced broadening of the HED in the initial $\sim 20~{\rm fs}$ is much larger than the broadening due to phonons.
We took into account both electron-electron and electron-phonon scattering during the whole time evolution, and showed that electron-phonon scattering does indeed play a minor role also in the initial, Coulomb-scattering-dominated, stage of the dynamics.

Intra- and inter-band coherences were previously considered~\cite{butscher_apl_2007,malic_prb_2011,kim_prb_2011,sun_prb_2012} in the framework of the density-matrix formalism (``semiconductor Bloch equations").
In the range of parameters we used here, the solution of the semiclassical Boltzmann equation agrees with these results.
Taking coherences into consideration gives minor oscillations of the electron density (analogous to the Rabi oscillations in a two-level system~\cite{Barnett_and_Radmore}), which damp out quickly due to dephasing induced by electron and phonon scattering.

Genuine quantum kinetic effects~\cite{HaugJauhoBook} affect the dynamics on time scales of a few fs.
To the best of our knowledge, quantum kinetic theory was never applied to ultrafast electron dynamics in graphene.
Although quantum kinetic effects are fundamental to describe the \emph{coherent buildup} of screening~\cite{HaugJauhoBook}, here we targeted the role of screening in the time window $20~{\rm fs} \lesssim t \lesssim 100~{\rm fs}$.
The effects of screening on this time scale are a dominant contribution to the dynamics, and it is unlikely that a more precise description of the buildup of screening can substantially alter this picture.
Indeed, it is known~\cite{HaugJauhoBook} that quasi-classical theories can be used to fit Coulomb quantum kinetics on a timescale longer than $20~{\rm fs}$.
We also point out that it is an extremely difficult task to estimate {\it a priori} the short initial transient in which the coherent buildup of screening takes place.
In an equilibrium state, basic consideration of the screening dynamics~\cite{Giuliani_and_Vignale} suggests that the buildup of screening, on a given length scale, should take place on a time scale comparable with the period of the plasma oscillations at the corresponding wavelength.
This implies that coherent screening buildup is faster in systems with a large carrier density (several hundreds ${\rm meV}$), which corresponds to the parameters that we use in this work.
However, the initial state that we consider is strongly displaced from equilibrium and the relation between coherent buildup of screening and carrier density is less clear.
It is arguable that electron-electron scattering is much more effective in a non-equilibrium state than in a thermal state, thus the electron-scattering-induced dephasing further reduces the time span where coherent buildup of screening plays a relevant role.

A second outcome of quantum kinetic theory is the ability to describe memory (or non-Markovian) effects, which could play in principle an important role on ultrafast time scales.
As discussed in Ref.~\onlinecite{snoke_annphys_2012}, the general effect of using a quantum kinetic equation is to introduce oscillations in the response of the system on very short timescales, typically of the order of ${\rm fs}$ in solid state systems.
We focused here on the time window $20~{\rm fs} \lesssim  t \lesssim 100~{\rm fs}$, where memory effects which survive for a few ${\rm fs}$ only are likely to be irrelevant.
Moreover, memory effects seem to be very important in systems with long-range Coulomb interactions~\cite{bonitz_jphys_1996}.
In our system, instead, electron-electron interactions are well screened since we are studying the dynamics of a high-density droplet of excited carriers.
In the case of effectively short-range interactions, memory effects are much less pronounced~\cite{bonitz_jphys_1996} and seem to be more important at low energies~\cite{bonitz_jphys_1996}.
In graphene the density-of-states vanishes at low energies, therefore further suppressing them.
Thus, although our theory does not apply down to the few-${\rm fs}$ time scale, where the coherent buildup of screening and memory effects take place, these effects do not change the dynamics in the window of interest, $20~{\rm fs} \lesssim t \lesssim 100~{\rm fs}$.
In this time window we demonstrated that it is of utmost importance to have an accurate description of screening, which dominates the relaxation dynamics.

An analytical treatment of the Coulomb collision integral distinguishes our work from previous ones~\cite{rana_prb_2007,malic_prb_2011,sun_prb_2012}.
In this respect, our main result, Eq.~(\ref{eq:kernelintegralsimplified}), expresses the Coulomb collision integral in terms of a compact and computationally-convenient 1d integral over the modulus $Q$ of the total momentum of a two-particle scattering process.
This approach allows us to carefully deal with {\it all} the singularities that arise in the limit of collinear electron-electron scattering.
Different screening models have been analyzed, and their impact on the collinear scattering singularities elucidated.
We also proposed a computationally-efficient way to take into account screening in Eqs.~(\ref{eq:lindhardTwoMu}), (\ref{eq:finalapproximation}), which is fully quantitative after thermalization occurs.

Solving numerically the semiclassical Boltzmann equation, we concluded that the particular form of screening one uses largely controls the speed of the relaxation dynamics in the sub-$100~{\rm fs}$ time range.
Different screening models yield markedly different time evolutions in a large portion of parameter space.

Our semi-analytical approach can be easily generalized to other carbon-based materials, such as bilayer and trilayer graphene and carbon nanotubes.

\begin{acknowledgments}
We thank Denis Basko, Annalisa Fasolino, Misha Katsnelson, Frank Koppens, Leonid Levitov, Allan MacDonald, Jairo Sinova, Justin Song, and Giovanni Vignale for very useful and stimulating discussions.
This work was supported by the Italian Ministry of Education, University, and Research (MIUR) through the program ``FIRB - Futuro in Ricerca 2010'' Grant No.~RBFR10M5BT (``PLASMOGRAPH: plasmons and terahertz devices in graphene''), ERC grants NANOPOT and STRATUS (ERC-2011-AdG No. 291198), EU grants RODIN and GENIUS, a Royal Society Wolfson Research Merit Award, EPSRC grants EP/K017144/1, EP/K01711X/1, EP/GO30480/1 and EP/G042357/1, and the Cambridge Nokia Research Centre.
We have made use of free software (www.gnu.org, www.python.org).
\end{acknowledgments}

\appendix

\section*{Appendix: details of numerical calculations}
\label{ssec:numerical}

The electron energy states are discretized on a uniform mesh $\{\varepsilon_{i}\}_{i=-L}^{L}$, centered around the Dirac point.
We take up to $L = 100$, with step $\varepsilon_{i+1} - \varepsilon_{i} = 25.0~{\rm meV}$.
The wave vectors of the phonon modes are discretized on a matching mesh $\{q_{i}\}_{i=1}^{L}$ with $q_{i} = \varepsilon_{i} / (\hbar v_{\rm F})$.
The phonon energies are approximated to a multiple of the energy step.
The total number of variables, including spin and valley degeneracy for the electrons and the four phononic modes, is then $4 \times 2 L + 4 \times L$.
The SBE are first-order differential equations which are solved using a standard fourth-order Runge-Kutta algorithm~\cite{NumericalRecipes}, with a time step $\delta t$ as small as $\delta t = 0.001~{\rm fs}$.

The kernels~(\ref{eq:PhononKernels}) of the e-ph interactions are computed at the beginning of the time evolution.
The largest computational burden is the evaluation of the Coulomb kernel~(\ref{eq:coulombkernel}), which scales cubically with the number of states $L$.
The Coulomb kernel is updated at a variable rate, depending on the stage of the time evolution.
For $t < 20.0~{\rm fs}$, when the system has not reached a thermal state yet, we update the kernel each $\delta t_{\rm C} = 2.0~{\rm fs}$.
We use $\delta t_{\rm C} = 5.0~{\rm fs}$ in the early cooling stage $t < 100.0~{\rm fs}$ and $\delta t_{\rm C} = 50.0~{\rm fs}$ afterwards.
We checked that the numerical results do not depend on the specific choice of these parameters.

Before updating the Coulomb kernel at time $t$, we estimate $T$ and $\mu_{s}$ of the electron population in the two bands.
To this end, we define the two functionals $\Phi_{1}[\varphi(\varepsilon)] \equiv \int_{0}^{\infty} d\varepsilon \varphi(\varepsilon)$ and $\Phi_{2}[\varphi(\varepsilon)] \equiv \int_{0}^{\infty} d\varepsilon \varphi(\varepsilon)^{2}$.
The values of the two functionals applied to the FD distribution can be computed exactly:
\begin{eqnarray}
\Phi_{1}[F(\varepsilon; \mu, T)] & = & - k_{\rm B} T \ln{F(\mu;0,T)} \nonumber \\
\Phi_{2}[F(\varepsilon; \mu, T)] & = & k_{\rm B} T \left [ F(\mu;0,T) - 1 \right ] + \mu \nonumber \\
& & - k_{\rm B} T \ln{F(-\mu;0,T)}~.
\end{eqnarray}
We tabulated these values on a mesh of $\mu$ and $T$.
Then, during the time evolution, we evaluate $\Phi_{1}[f_{\ell}(\varepsilon)]$ and $\Phi_{2}[f_{\ell}(\varepsilon)]$.
We then find the values $\mu$ and $T$ on the mesh which minimize $\sum_{i\in\{1,2\}}|\Phi_{i}[f_{\ell}(\varepsilon)] - \Phi_{i}[F(\varepsilon; \mu, T)]|$.
This procedure yields our estimate for $T$ and $\mu_{\rm c} = \mu$, $\mu_{\rm v} = -\mu$, with the advantage that can be applied automatically during the time evolution and is more robust than a standard fitting procedure.
The consistency of the estimate for $\mu_{\rm s}$ and $T$ is checked at the end of the time evolution using a more precise fitting procedure for the electron distribution.

The values of $\mu_{s}$ and $T$ are used to compute $\chi^{(0)}(q,\omega;T)$, according to Eq.~(\ref{eq:lindhardTwoMu}).
The polarization function is evaluated on a 2d mesh for the modulus of the transferred wave vector $q$ [with step $\delta q \simeq 10.0~{\rm meV}$ and maximum value $q_{\rm max} = 2.0 \times \varepsilon_{L} / (\hbar v_{\rm F})$] and frequency $\omega$ (symmetric about $\omega = 0$, with $\omega_{\rm max} = 1.7 \times v_{\rm F} q_{\rm max}$).
The integration over $\varepsilon'$ in Eq.~(\ref{eq:lindhardTwoMu}) is performed over a rather rough mesh with $20$ points up to $10.0~{\rm eV}$ using rectangles rule.
This is sufficient to capture the effects of $T$ to good accuracy, as we tested by reproducing with this method the Lindhard function at equilibrium~\cite{RVAPM2009}.
The procedure of regularization of the polarization function introduced in Sec.~\ref{ssec:screeningModels} is implemented by replacing $\chi^{(0)}(q,\omega;T)$ with $\chi^{(0)}(q, v_{\rm F}q - \Lambda_{\rm E} / \hbar;T)$ when $\hbar \omega \in [\hbar v_{\rm F} q - \Lambda_{\rm E},\hbar v_{\rm F} q]$ below the light cone (and similarly above the light cone).
Special care has to be taken when $\hbar v_{\rm F} q < \Lambda_{\rm E}$ in the proximity of the origin.
In this Article we used $\Lambda_{\rm E} = 20.0~{\rm meV}$.

Finally, the Coulomb kernel is computed according to Eqs.~(\ref{eq:kernelintegralsimplified}) and (\ref{eq:AugerResult}).
For each choice of the energies of the incoming and outgoing particles, and the modulus $Q$ of the total momentum (only one value is possible for Auger processes) the closer values for the transferred wave vector and frequency are matched on the mesh for the polarization function.
For intra-band and inter-band terms, the integral over the total momentum is performed using the standard Simpson rule~\cite{AbramowitzStegun} mesh with $21$ points.
Before performing the integration, a change of variables is performed to an effective angle variable $\phi$ given by $Q = (Q_{\rm max} + Q_{\rm min}) / 2 + [(Q_{\rm max} - Q_{\rm min})/2] \cos\phi$, to improve the precision of the integral at the extremes.
To numerically avoid the collinear divergence we restrict the integration variable in the interval $\phi \in [\delta\phi,\pi - \delta\phi]$, where we take $\delta \phi = 0.0001$ (different choices over a few orders of magnitude do not contribute substantial changes to the results).

The speed in solving the SBE with this method can be increased substantially by reducing $L$, and the final results are qualitatively correct even with $L = 50$ and $\varepsilon_{i+1} - \varepsilon_{i} = 100.0~{\rm meV}$ although, in this case, the different phonon energies are not resolved.

\end{document}